\documentclass[pre,twocolumn,amsmath,amssymb,floatfix,superscriptaddress,showpacs]{revtex4}

\usepackage{graphicx}
\usepackage{latexsym}
\usepackage{amsmath}
\usepackage{amssymb}
\usepackage{amsfonts}
\usepackage{bm}

\newcommand{\la}{\left<}
\newcommand{\ra}{\right>}

\newcommand{\nvecl}{\underline{n}_l}
\newcommand{\pvec}{\ensuremath{\underline{p}}}

\newcommand{\rvec}{\ensuremath{\underline{r}}}
\newcommand{\rvecl}{\ensuremath{\underline{r}_l}}

\newcommand{\vvec}{\ensuremath{\underline{v}}}
\newcommand{\ddiff}{\ensuremath{\text{d}}}

\newcommand{\nlx}{n_{l,x}}
\newcommand{\nly}{n_{l,y}}

\newcommand{\rl}{r_l}

\newcommand{\rix}{r_{i,x}}
\newcommand{\riy}{r_{i,y}}

\newcommand{\vix}{v_{i,x}}
\newcommand{\viy}{v_{i,y}}

\newcommand{\nsp}{\mbox{$n_{\rm sp}$}}

\newcommand{\Ksp}{\mbox{$K_{\rm sp}$}}
\newcommand{\Unsp}{\mbox{$U_{\rm nsp}$}}
\newcommand{\Usp}{\mbox{$U_{\rm sp}$}}
\newcommand{\Rsp}{\mbox{$R_{\rm sp}$}}
\newcommand{\Nsp}{\mbox{$N_{\rm sp}$}}
\newcommand{\rhosp}{\mbox{$\rho_{\rm sp}$}}
\newcommand{\Kbb}{\mbox{$K_{\rm b}$}}
\newcommand{\Rbb}{\mbox{$R_{\rm b}$}}
\newcommand{\Ubb}{\mbox{$U_{\rm b}$}}
\newcommand{\Nbb}{\mbox{$N_{\rm b}$}}
\newcommand{\rc}{\mbox{$r_{\rm c}$}}

\newcommand{\kB}{\mbox{$k_{\rm B}$}}

\newcommand{\NVgT}{\ensuremath{\text{NV}\gamma\text{T}}}
\newcommand{\NVtT}{\ensuremath{\text{NV}\tau\text{T}}}

\newcommand{\muA}{\ensuremath{\mu_\mathrm{A}}}

\newcommand{\muAhat}{\ensuremath{\hat{\mu}_\mathrm{A}}}
\newcommand{\muAidhat}{\ensuremath{\hat{\mu}_\mathrm{A,id}}}
\newcommand{\muAexhat}{\ensuremath{\hat{\mu}_\mathrm{A,ex}}}
\newcommand{\muF}{\ensuremath{\mu_\mathrm{F}}}
\newcommand{\muFtild}{\ensuremath{\tilde{\mu}_\mathrm{F}}}
\newcommand{\muFstar}{\ensuremath{\mu_{\star}}}

\newcommand{\Cttild}{\ensuremath{\tilde{c}(t)}}
\newcommand{\Ctild}{\ensuremath{\tilde{c}}}

\newcommand{\Gpr}{\delta G}
\newcommand{\Geq}{G_\mathrm{eq}}
\newcommand{\GF}{G_\mathrm{F}}

\newcommand{\tauhat}{\hat{\tau}}

\newcommand{\tauidhat}{\hat{\tau}_\mathrm{id}}
\newcommand{\tauexhat}{\hat{\tau}_\mathrm{ex}}

\newcommand{\Phat}{\hat{P}}
\newcommand{\Ocal}{\ensuremath{{o}}}
\newcommand{\Ocalhat}{\ensuremath{\hat{o}}}

\newcommand{\ahat}{\ensuremath{\hat{a}}}

\newcommand{\Hhat}{\hat{\cal H}}
\newcommand{\Hidhat}{\hat{\cal H}_\mathrm{id}}
\newcommand{\Hexhat}{\hat{\cal H}_\mathrm{ex}}

\newcommand{\dtMD}{\delta t_\mathrm{MD}}

\newcommand{\tauA}{t_\mathrm{A}}
\newcommand{\xAf}{x_\mathrm{A}(f)}
\newcommand{\tstar}{t_{\star}}
\newcommand{\Gstar}{G_{\star}}

\newcommand{\tsamp}{\Delta t}
\newcommand{\xsamp}{\Delta x}
\newcommand{\ttraj}{t_\mathrm{traj}}
\newcommand{\ttemp}{t_\mathrm{temp}}
\newcommand{\mbound}{m_\mathrm{min}}

\newcommand{\Tver}{\ensuremath{{\cal P}_{\tsamp}}}

\newcommand{\fDebye}{f_\mathrm{Debye}}

\newcommand{\peq}{p_\mathrm{eq}}

\newcommand{\state}{\sigma}
\newcommand{\statep}{\sigma'}

\begin{document}

\title{Shear-stress fluctuations in self-assembled transient elastic networks}

\author{J.P.~Wittmer}
\email{joachim.wittmer@ics-cnrs.unistra.fr}
\affiliation{Institut Charles Sadron, Universit\'e de Strasbourg \& CNRS, 23 rue du Loess, 67034 Strasbourg Cedex, France}
\author{I. Kriuchevskyi}
\affiliation{Institut Charles Sadron, Universit\'e de Strasbourg \& CNRS, 23 rue du Loess, 67034 Strasbourg Cedex, France}
\author{A. Cavallo}
\affiliation{Institut Charles Sadron, Universit\'e de Strasbourg \& CNRS, 23 rue du Loess, 67034 Strasbourg Cedex, France}
\author{H.~Xu}
\affiliation{LCP-A2MC, Institut Jean Barriol, Universit\'e de Lorraine \& CNRS, 1 bd Arago, 57078 Metz Cedex 03, France}
\author{J. Baschnagel}
\affiliation{Institut Charles Sadron, Universit\'e de Strasbourg \& CNRS, 23 rue du Loess, 67034 Strasbourg Cedex, France}

\begin{abstract}
Focusing on shear-stress fluctuations we investigate numerically a simple generic model for 
self-assembled transient networks formed by repulsive beads reversibly bridged by ideal springs.
With $\tsamp$ being the sampling time and $\tstar(f) \sim 1/f$ the Maxwell relaxation time
(set by the spring recombination frequency $f$) the dimensionless parameter 
$\xsamp = \tsamp/\tstar(f)$ is systematically scanned from the liquid limit ($\xsamp \gg 1)$
to the solid limit ($\xsamp \ll 1$) where the network topology is quenched and an ensemble 
average over $m$ independent configurations is required. 
Generalizing previous work on permanent networks it is shown that the shear-stress relaxation 
modulus $G(t)$ may be efficiently determined for all $\xsamp$ using the simple-average 
expression $G(t) = \muA - h(t)$ with $\muA = G(0)$ characterizing the canonical-affine 
shear transformation of the system at $t=0$ and $h(t)$ the (rescaled) mean-square displacement 
of the instantaneous shear stress as a function of time $t$.  
This relation is compared to the standard expression $G(t) = \Cttild$ using the (rescaled) 
shear-stress autocorrelation function $\Cttild$.  
Lower bounds for the $m$ configurations required by both relations are given.
\end{abstract}
\date{\today}
\maketitle

\section{Introduction}
\label{sec_intro}

\begin{figure}[t]
\centerline{\resizebox{0.9\columnwidth}{!}{\includegraphics*{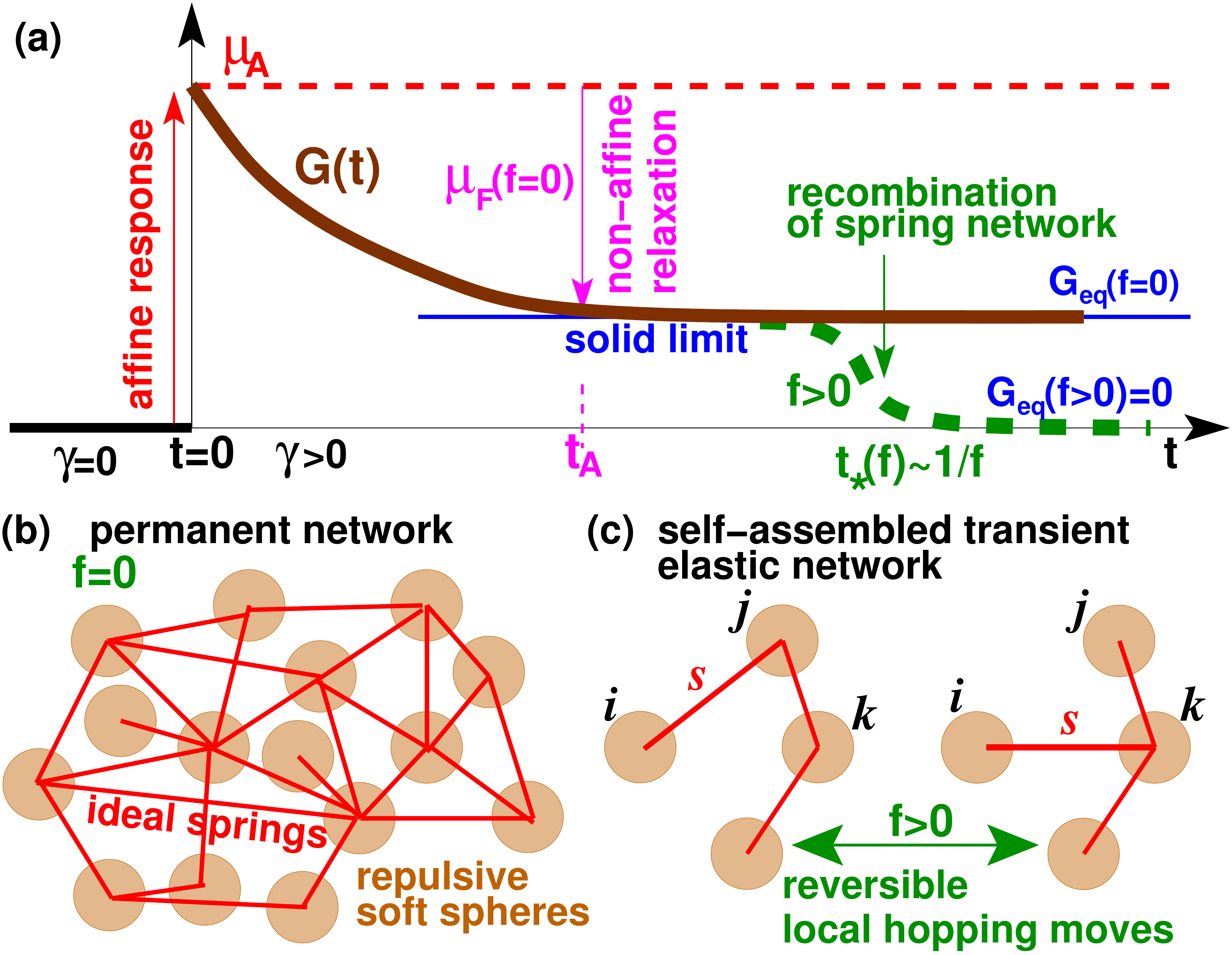}}}
\caption{Addressed problem:
{\bf (a)}
Shear-stress relaxation modulus $G(t)$ after a tiny step strain 
$\delta \gamma$ is imposed at $t=0$ (bold lines).  
{\bf (b)} 
Permanent elastic network formed by beads connected by ideal harmonic springs 
(thin solid lines) without recombinations ($f=0$).
{\bf (c)}
Self-assembled transient elastic network created by reversibly
breaking and recombining springs with an attempt frequency $f > 0$ per spring
subject to a Metropolis criterion. The spring $s$ thus connects the beads $i$ and $j$
on the left and the beads $i$ and $k$ on the right. 
\label{fig_sketch}
}
\end{figure}

\subsection{Background: Permanent networks}
\label{intro_back}

A central rheological property characterizing the linear shear-stress response in
isotropic amorphous solids and glasses \cite{Alexander98,HansenBook,GoetzeBook} and 
visco-elastic fluids \cite{DegennesBook,DoiEdwardsBook,WittenPincusBook,RubinsteinBook} 
is the shear relaxation modulus $G(t)$ sketched in panel (a) of Fig.~\ref{fig_sketch}.
Experimentally, $G(t) = \delta \tau(t)/\delta \gamma$ may be obtained from the average
stress increment $\delta \tau(t)$ as a function of time $t$ after a small step strain 
$\delta \gamma$ has been imposed at $t=0$. 
As indicated by the solid horizontal line in panel (a), $G(t)$ yields the equilibrium shear modulus 
$\Geq$ of the system in the long-time limit for $t \gg \tstar$ with $\tstar$ being the terminal 
stress relaxation time \cite{RubinsteinBook,DoiEdwardsBook}. 
Focusing on permanent elastic networks above the percolation threshold 
\cite{DegennesBook,StaufferBook,Zippelius06} with a {\em finite} shear modulus $\Geq$, 
as sketched in panel (b) of Fig.~\ref{fig_sketch}, it has been shown  \cite{WXB16} 
that $G(t)$ may be determined conveniently in computer simulations using the ``simple average" 
expression 
\begin{equation}
G(t) = \muA - h(t)
\label{eq_key}
\end{equation}
with $\muA = G(0)$ being the ``affine shear elasticity" characterizing the canonical-affine 
shear transformation (Appendix~\ref{theo_affine}) of the system at $t=0$ 
\cite{WXP13,WXB15,WXBB15,WKB15,WXB16} and
$h(t) = \beta V/2 \ \langle (\tauhat(t)-\tauhat(0))^2 \rangle$ the
(rescaled) mean-square displacement (MSD) of the instantaneous shear stress $\tauhat(t)$.
Here $\beta=1/T$ stands for the inverse temperature and $V$ for the volume of the simulation box.
See Appendix~\ref{theo_muAhat} for the related definitions of the instantaneous shear stress $\tauhat$
and the instantaneous affine shear elasticity $\muAhat$.
Interestingly, the expectation value of Eq.~(\ref{eq_key}) does not depend on the sampling time 
$\tsamp$ even if much smaller times than the terminal time $\tstar$ are probed \cite{WXB16}.
For sufficiently large systems Eq.~(\ref{eq_key}) can be demonstrated using the simple-average 
transformation behavior \cite{AllenTildesleyBook,Lebowitz67} of $\muA$ 
and $h(t)$ between the $\NVgT$-ensemble at constant particle number $N$, volume $V$, 
shear strain $\gamma$ and temperature $T$ and the conjugated $\NVtT$-ensemble at an 
imposed average shear stress $\tau$ \cite{WXB16}.

Albeit the equilibrium shear modulus $\Geq$ may in principal be determined from the
long-time limit of Eq.~(\ref{eq_key}), most numerical studies
\cite{Barrat88,WTBL02,TWLB02,Szamel15,XWP12,WXP13,WXB15,WXBB15,WKB15,WXB16}
use instead the stress-fluctuation formula $\Geq = \GF$ with
\begin{eqnarray}
\GF & \equiv & \muA - \muF = (\muA-\muFtild) + \muFstar \label{eq_GF} \\
\mbox{and } \muF & \equiv & \beta V \langle \delta \tauhat^2 \rangle = \muFtild - \muFstar
\label{eq_muF}
\end{eqnarray}
standing for the rescaled shear-stress fluctuations. We have introduced here for later 
convenience the two terms $\muFtild \equiv \beta V \langle \tauhat^2 \rangle$
and $\muFstar \equiv \beta V \langle \tauhat \rangle^2$.
As sketched in panel (a) of Fig.~\ref{fig_sketch}, $\muF$ corresponds to the (free) energy relaxed by
non-affine displacements after an initial canonical-affine shear strain $\delta \gamma$ is imposed.
Note that $\GF$ is a special case of the general stress fluctuation relations for elastic moduli
\cite{Hoover69,Barrat88,Lutsko89,Barrat13}. 
As stressed elsewhere \cite{WXP13,WXB15,WXB16}, being ``fluctuations" (not ``simple averages")
\cite{AllenTildesleyBook,WXB16} the expectation values of $\GF$, $\muF$ and $\muFstar$ may depend 
strongly on the sampling time $\tsamp$ (as often marked below by indicating $\tsamp$ as additional argument)
and converge very slowly to their asymptotic static limit for $\tsamp \gg \tstar$. 
This behavior is not due to aging or equilibration problems but simply caused by the 
finite time needed for the stress fluctuations to explore the phase space \cite{WXP13}. 
Interestingly, using Eq.~(\ref{eq_key}) and assuming time translational invariance 
it can be shown that $\GF(\tsamp)$ and $G(t)$ are related by
\begin{equation}
\GF(\tsamp) = \frac{2}{\tsamp^2} \int_0^{\tsamp} (\tsamp -t) \ G(t) \ \ddiff t,
\label{eq_GF_Gt}
\end{equation}
i.e. $\GF(\tsamp)$ is a (weighted) average of $G(t)$ \cite{foot_polymer,foot_simrel}.
It converges thus more slowly to $\Geq$ but this with a better statistics.
See Ref.~\cite{WXB15} and Appendix~\ref{simu_conseq} for details.

\subsection{New focus: Transient self-assembled networks}
\label{intro_new}
We generalize here our previous work on solid bodies \cite{WXB15,WXB16}
to visco-elastic liquids \cite{DegennesBook,WittenPincusBook,RubinsteinBook}.
The {\em first goal} of the present work is to introduce and to characterize numerically
a simple model for transient self-assembled networks
\cite{Porte03,Safran06,Ligoure08,Leibler11,Leibler13,Kob13,Friedrich10}.
As sketched in panel (c) of Fig.~\ref{fig_sketch}, repulsive ``harmonic spheres"
\cite{Berthier10,Berthier11a} are reversibly bridged by ideal springs.
It is assumed that the springs break and recombine locally with a Monte Carlo (MC) hopping
frequency $f$ in a similar manner as in earlier work on equilibrium polymer systems \cite{WMC98b,HXCWR06}.
As sketched by the bold dashed line in panel (a) of Fig.~\ref{fig_sketch},
these transient networks are shown to be simple Maxwell fluids \cite{RubinsteinBook},
i.e. the shear-stress relaxation modulus decays exponentially
\begin{equation}
G(t) \approx \Gstar \exp(-x) \mbox{ with } x \equiv t/\tstar(f) \mbox{ for } t/\tauA \gg 1 
\label{eq_Maxwell}
\end{equation}
with $\tauA$ being a local time scale characterizing the decay of the initial affine displacements, 
$\tstar(f) \sim 1/f$ the Maxwell time and $\Gstar$ the intermediate plateau modulus 
set by the equilibrium shear modulus $\Geq$ for permanent springs ($f=0)$.
From the rheological point of view our model is very similar to patchy colloids 
\cite{Leibler13,Kob13} or ``vitrimers" \cite{Leibler11}, 
i.e. covalent polymer networks that can rearrange their topology via a bond 
shuffling mechanism.
Rheologically similar self-assembled transient networks may also be formed by hyperbranched 
polymer chains with sticky end-groups \cite{Friedrich10} or microemulsions bridged by 
telechelic polymers \cite{Porte03,Safran06,Ligoure08}.
While mainly keeping the sampling time $\tsamp$ constant,
we systematically scan the dimensionless attempt frequency 
$\xsamp \equiv \tsamp/\tstar(f) \sim f$
from the liquid state ($\xsamp \gg 1$), where the network topology is annealed,
down to the solid limit ($\xsamp \ll 1$), where the recombination events become irrelevant
and the particle permutation symmetry is lifted \cite{WittenPincusBook}.
Due to detailed balance this is done while keeping unchanged all static properties
related to pair correlations. The $\xsamp$-dependence reported below for $G(t)$ or $\GF$
thus cannot be traced back to pair correlations as often assumed for glass-forming systems 
\cite{HansenBook,GoetzeBook}.
By integration of the general relation Eq.~(\ref{eq_GF_Gt}) for a Maxwell fluid, 
Eq.~(\ref{eq_Maxwell}), one expects in fact the shear-stress fluctuations to be given by
\begin{equation}
\GF(\xsamp) \equiv \muA - \muF(\xsamp)  =   \Gstar \ \fDebye(\xsamp) \label{eq_GFfDebye}
\end{equation}
with $\fDebye(x) = 2(\exp(-x)-1+x)/x^2$ being the Debye function well-known in polymer physics 
\cite{DoiEdwardsBook,RubinsteinBook}.
We shall see that this important relation allows to interpolate our numerical data 
between the solid limit, where $\GF(\xsamp) \to \Gstar$ and $\muF(\xsamp) \to \muA-\Gstar$ 
for $\xsamp \ll 1$, and the liquid limit, where $\GF(\xsamp) \to 0$ and 
$\muF(\xsamp) \to \muA$ for $\xsamp \gg 1$.

Using our simple model the {\em second goal} of this study is to show 
that Eq.~(\ref{eq_key}) does not only hold for elastic solids ($\xsamp \ll 1$) 
but more generally for visco-elastic bodies, i.e. for all values of $\xsamp$. 
We shall compare this relation to the widely assumed expression 
\cite{AllenTildesleyBook,HansenBook,DKG91,Klix12,Szamel15,Leibler13}
\begin{equation}
G(t) = \Ctild(t) \mbox{ with } \Cttild \equiv \beta V \la \tauhat(t) \tauhat(0) \ra
\label{eq_keyapprox}
\end{equation}
being the (rescaled) shear-stress autocorrelation function (ACF).
Albeit Eq.~(\ref{eq_keyapprox}) is incorrect for general elastic bodies 
\cite{WXB15,WXBB15,WKB15,WXB16}, it may be justified under the condition
\begin{equation}
\muA \stackrel{!}{=} \tilde{c}(t=0) \equiv \muFtild.
\label{eq_condition}
\end{equation}
While this condition indeed holds {\em on average} for self-assembled networks, 
it requires on the numerical side that either $\xsamp \gg 1$, or, equivalently, an 
ensemble-average over a large number $m$ of independent configurations. 
Being thus both in principle acceptable means to determine $G(t)$ for any $\xsamp$,
this does, of course, not imply that Eq.~(\ref{eq_key}) and Eq.~(\ref{eq_keyapprox})
have the same statistics. We shall thus attempt to characterize the standard deviations 
of both relations and estimate lower bounds for the number of configurations required.

\subsection{Outline}
\label{intro_outline}

Our numerical model is formulated in Sec.~\ref{sec_algo} where we also address several 
technical questions. Our central numerical findings are then discussed in Sec.~\ref{sec_simu}.
Carefully stating the subsequent time and ensemble averages performed, we present in 
Sec.~\ref{simu_static} the pertinent static and quasi-static properties. 
The MSD $h(t)$ is described in Sec.~\ref{simu_MSD} where we test Eq.~(\ref{eq_key}) 
numerically by comparing it to the shear response modulus $G(t)$ obtained by applying 
explicitly a small step strain $\delta \gamma$.
For the available $m=100$ configurations Eq.~(\ref{eq_keyapprox}) is shown in Sec.~\ref{simu_ACF}
to be a poor approximation of $G(t)$ for $\xsamp \ll 1$.
The number of configurations required by, respectively, Eq.~(\ref{eq_key}) and 
Eq.~(\ref{eq_keyapprox}) are estimated in Sec.~\ref{simu_mbound}.
Section~\ref{sec_conc} contains a summary of the present work and an outline of open questions.
Less central issues are regrouped in the Appendix.
Concepts and definitions already stated elsewhere \cite{WXP13,WXB15,WXBB15,WKB15,WXB16}
are reminded in Appendix~\ref{theo_affine} and Appendix~\ref{theo_muAhat}.
The theoretical derivations of Eq.~(\ref{eq_GF}) and Eq.~(\ref{eq_key}) can be found
in Appendix~\ref{theo_GF} and Appendix~\ref{theo_Gt}. 
Being not based on the transformation behavior between conjugated ensembles used in
our previous work \cite{WXBB15,WKB15,WXB16}), these direct demonstrations are relevant
for (complex) liquid systems with vanishing equilibrium shear modulus the present
study focuses on.
Computational results related to the sampling time $\tsamp$ are briefly
discussed in Appendix~\ref{simu_dt} and Appendix~\ref{simu_conseq}.

\begin{figure}[t]
\centerline{\resizebox{0.9\columnwidth}{!}{\includegraphics*{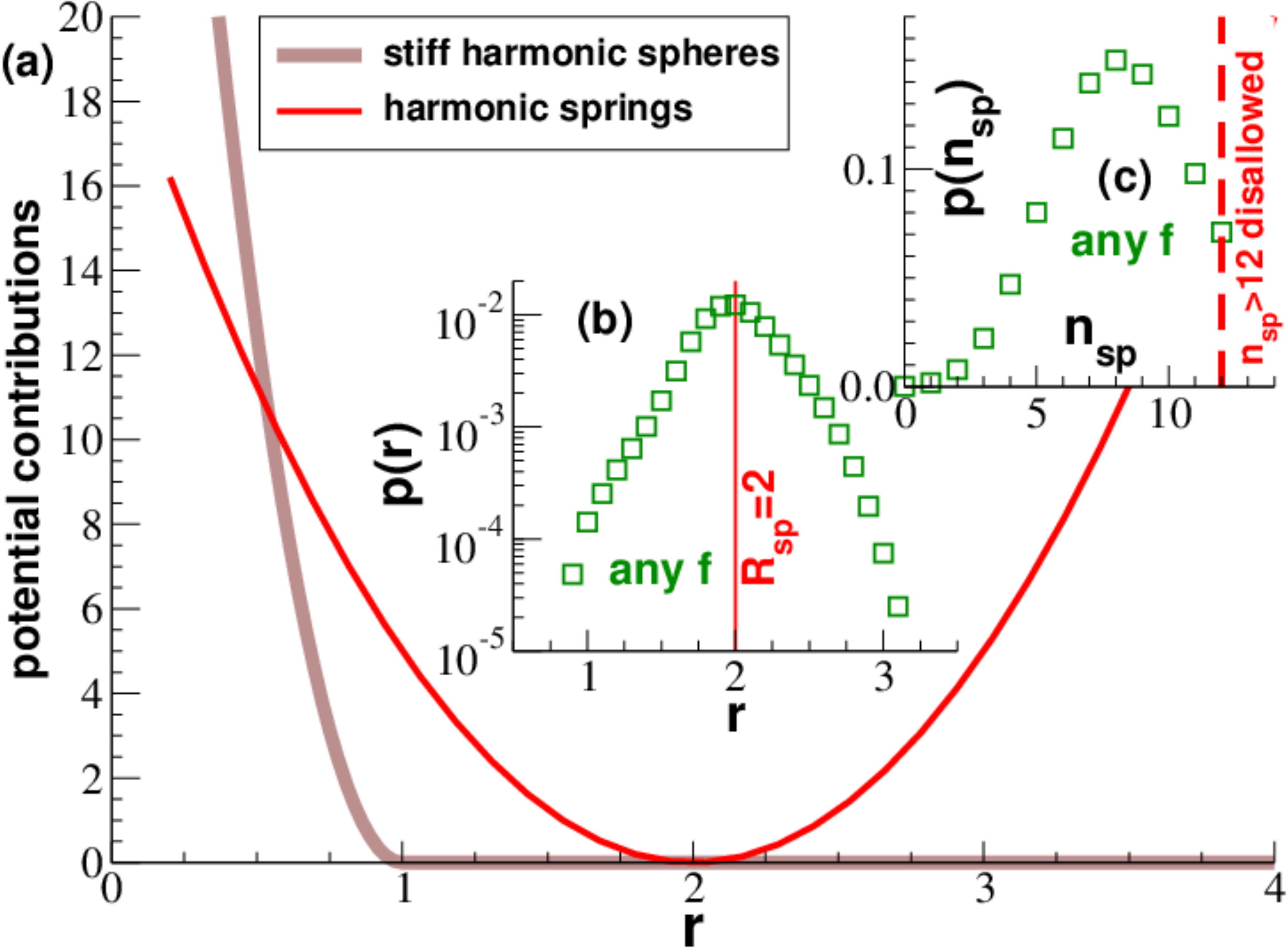}}}
\caption{(Color online)
Some technical details:
{\bf (a)}
Model Hamiltonian with the bold line indicating the purely repulsive interaction between 
``harmonics spheres" \cite{Berthier10} and the thin line the ideal spring between connected beads,
{\bf (b)} 
distribution $p(r)$ of spring lengths $r$ showing a maximum around
the minimum of the spring potential at $\Rsp=2$ and
{\bf (c)}
distribution $p(\nsp)$ of the number of springs $\nsp$ being connected to a bead
showing a maximum at $\nsp \approx 8$. Only a negligible number of beads
is not connected ($\nsp=0$) or are dangling ends ($\nsp=1$).
In the current work $\nsp \le 12$ is imposed.
The distributions shown in panel {\bf (b)} and {\bf (c)} are identical for all 
attempt frequencies $f$ due to detailed balance.
\label{fig_algo_pot}
}
\end{figure}

\begin{figure}[t]
\centerline{\resizebox{0.9\columnwidth}{!}{\includegraphics*{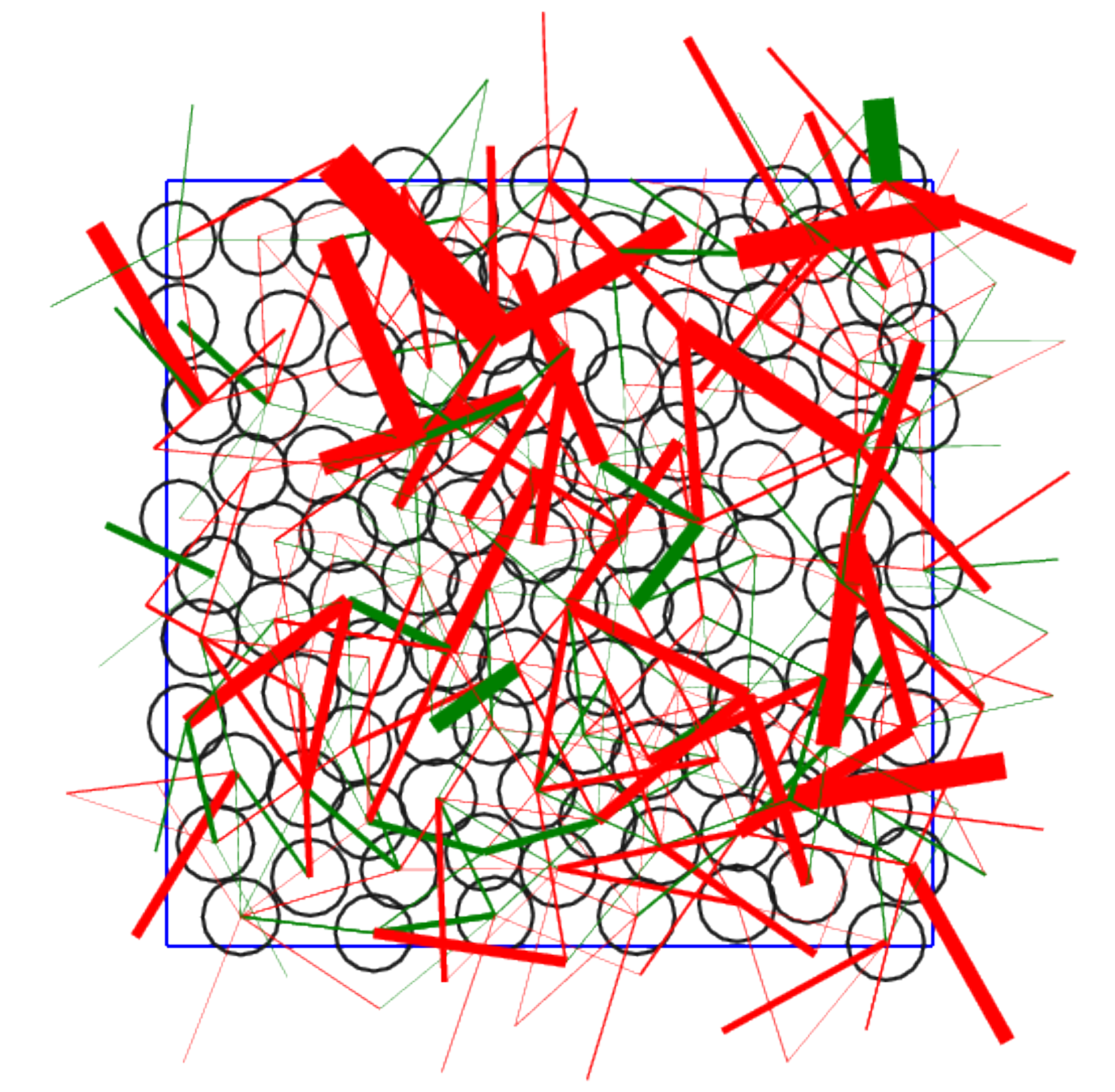}}}
\caption{(Color online)
Snapshot of small square subvolume of linear length $10$ containing $103$ beads (disks)
connected by $407$ springs (straight lines). The width of the spring lines is
proportional to the energy of the spring potential, Eq.~(\ref{eq_Uspring}).
Short springs with $r < \Rsp=2$ repel the beads (green lines),
longer springs (red lines) keep them together.
\label{fig_algo_snap}
}
\end{figure}

\section{Algorithm and technical details}
\label{sec_algo}

As sketched in Fig.~\ref{fig_sketch} we use a generic model for self-assembled 
elastic networks in $d=2$ dimensions where beads are reversibly bridged by ideal springs.
These springs recombine locally with a Monte Carlo (MC) attempt frequency $f$.
Lennard-Jones (LJ) units are used throughout this work \cite{AllenTildesleyBook} and 
the particle mass $m$, Boltzmann's constant $\kB$ and the temperature $T=1/\beta$ are set to unity. 
Periodic simulation boxes of constant volume $V=L^d$ and linear box size $L=100$ are used.
A standard Euclidean metric with a shear strain $\gamma=0$ can be assumed (square box) 
if not stated otherwise.
Moreover, the total number $\Nbb$ of beads and the number $\Nsp$ of 
springs are kept constant in the present work.

\begin{figure}[t]
\centerline{\resizebox{0.9\columnwidth}{!}{\includegraphics*{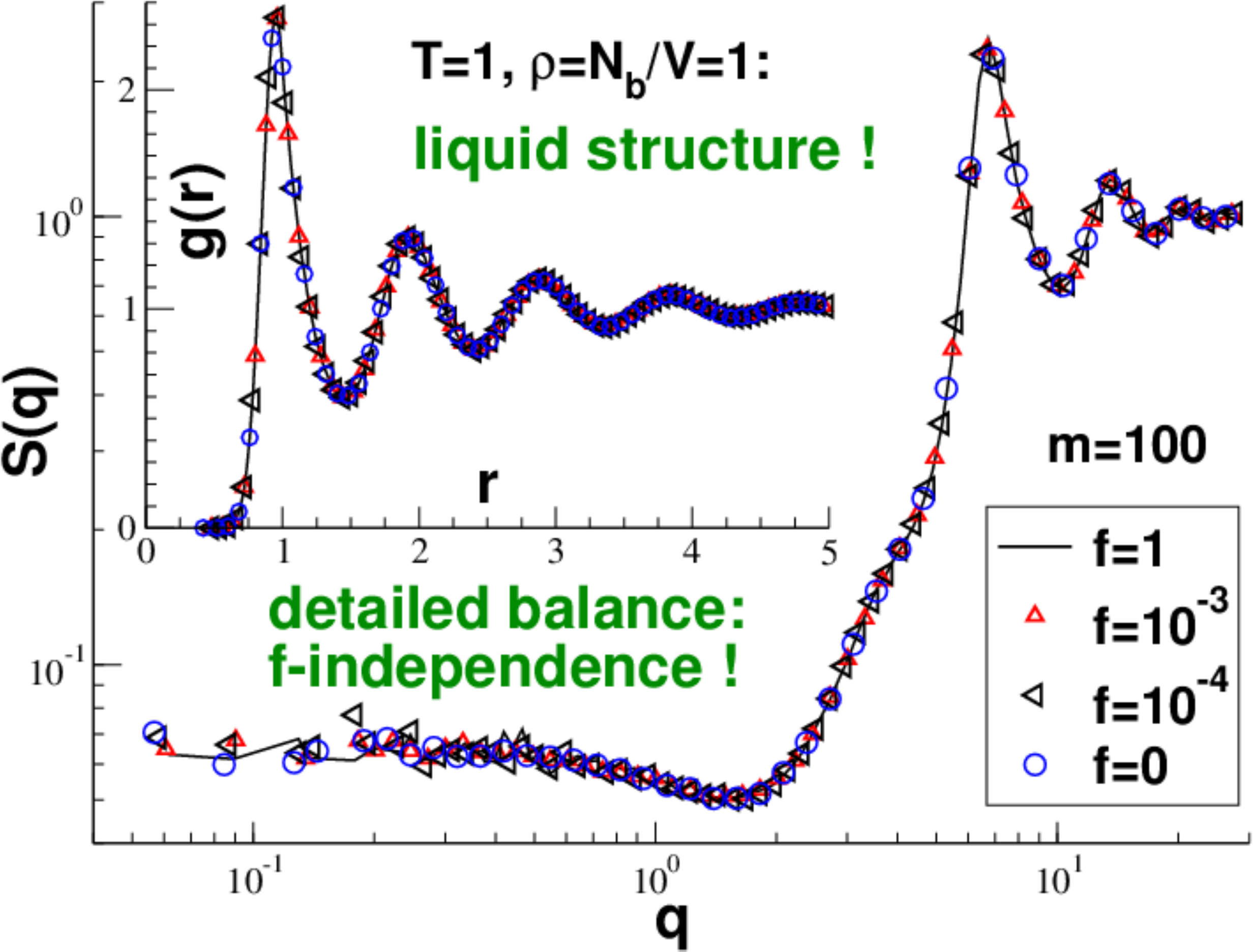}}}
\caption{(Color online)
Pair correlations for several attempt frequencies $f$.
Inset: Radial pair correlation distribution function $g(r)$ with $r$ being the distance
between two beads \cite{HansenBook}.
Main panel: Total coherent structure function $S(q)$ with $q$ being the length of the wavevector.
\label{fig_pairV}
}
\end{figure}

As shown by the bold solid line in panel (a) of Fig.~\ref{fig_algo_pot}, the particles are 
modeled as ``harmonic spheres" \cite{Berthier10} interacting through the purely repulsive potential 
\begin{equation}
\Ubb(r) = \frac{\Kbb}{2} \left(r - \Rbb \right)^2 \mbox{ for } r \le \Rbb
\label{eq_Ubead}
\end{equation}
and $\Ubb(r)=0$ elsewhere. The minimum of the shifted harmonic potential is used
as cut-off to avoid truncation effects and impulsive corrections for the determination
of the affine shear elasticity $\muA$ as described in Ref.~\cite{XWP12}.
The bead diameter is arbitrarily set to unity, $\Rbb=1$, and a rather stiff spring constant $\Kbb=100$ 
is used making the beads very repulsive. The simulation box contains $\Nbb=10^4$ beads, 
i.e. the number density $\rho=\Nbb/V$ of the beads is set to unity. 
Due to the strong repulsion and the high number density, the bead distribution is always 
macroscopically homogeneous and the overall density fluctuations are weak. 
This has been checked using snapshots, as the one shown in Fig.~\ref{fig_algo_snap},
and the standard radial pair correlation function $g(r)$ 
and its Fourier transform $S(q)$ \cite{AllenTildesleyBook} 
as presented in Fig.~\ref{fig_pairV}. 

The bonding of two beads is described by 
\begin{equation}
\Usp(r) = \frac{\Ksp}{2} \left(r - \Rsp \right)^2 
\label{eq_Uspring}
\end{equation}
with $\Rsp=2$ and $\Ksp=10$ as shown by the thin line in panel (a) of Fig.~\ref{fig_algo_pot}. 
Note that the minimum $\Rsp$ of the bonding potential is much larger than the bead diameter $\Rbb$.
There is thus no repulsion between two beads at $r \approx \Rsp$ and no sudden acceleration is
felt (on average) if a bond is broken. As seen in panel (b) of Fig.~\ref{fig_algo_pot}, 
the probability distribution $p(r)$ of springs of length $r$ has a sharp maximum at $\Rsp$
and the number of springs with $r < 1$ or $r > 3$ is negligible.
Our box contains a constant number $\Nsp = 4 \Nbb$ of springs, i.e. on average a bead
is connected by $\nsp=8$ springs. This corresponds roughly to the maximum of the distribution 
$p(\nsp)$ of the number $\nsp$ of springs connected to a given bead presented in panel (c)
of Fig.~\ref{fig_algo_pot}. Since there is no direct interaction (repulsion) between the
springs, the maximum number of springs connected to a bead is limited to $\nsp=12$ \cite{foot_nsp}.

As sketched in panel (c) of Fig.~\ref{fig_sketch}, the network is reorganized by attempting 
with a frequency $f$ local hopping moves for each spring. This is done by choosing first randomly 
a spring $s$ connecting two beads $i$ and $j$. If the spring length $r$ is smaller than a cut-off 
radius $\rc=5$, the connection to one bead is broken, say bead $j$, and we attempt to reconnect 
the spring to a randomly chosen monomer $k$ (different from $i$ or $j$) taken randomly from a 
neighbor list of beads with distance $r < \rc$ from the pivot monomer $i$ and having less than 
$\nsp=12$ springs attached \cite{foot_DB}. 
Using the energy change due to the different lengths of the suggested and the original spring state,
the move is accepted subjected to a standard Metropolis acceptance criterion 
\cite{AllenTildesleyBook,LandauBinderBook}.
The parameter $\rc$ is chosen sufficiently small to reduce the neighbor list and to yield a reasonable, 
not too small acceptance rate $A \approx 0.1$ (found to be identical for all $f$).
The computational load required by the reorganization of the network topology becomes 
negligible below an attempt frequency $f=0.01$.  

\begin{table}[t]
\begin{tabular}{|c|c||c|c|c|c|c|c|}
\hline \hline
$f$      &
$\xsamp$ & 
$e$      & 
$P$      & 
$\muA$   & 
$\muFtild$& 
$\muFstar$ &
$\GF$ 
\\ \hline
1.0  &6250    & 2.442 & 1.73 & 33.1 & $33      $& $\approx 0$& $ \approx 0$ \\
0.1  &625     & 2.442 & 1.73 & 33.1 & $33 (0.1)$& $\approx 0$& $ \approx 0$\\
0.01 &62.5    & 2.442 & 1.72 & 33.1 & $33 (0.3)$& $0.5 (0.1)$& $0.5  (0.3)$\\
E-03 &6.25    & 2.443 & 1.73 & 33.1 & $36 (1.1)$& $5   (0.7)$& $2.9  (0.8)$\\
E-04 &0.625   & 2.442 & 1.73 & 33.2 & $32 (1.9)$& $13  (1.9)$& $13.6 (0.4)$\\
E-05 &0.0625  & 2.441 & 1.74 & 33.2 & $29 (1.8)$& $14  (1.8)$& $17.4 (0.1)$\\
E-06 &6.25E-03& 2.443 & 1.73 & 33.3 & $35 (2.7)$& $20  (2.7)$& $17.8      $\\
E-07 &6.25E-04& 2.442 & 1.73 & 33.2 & $32 (2.5)$& $18  (2.5)$& $17.9      $\\
0    &0       & 2.440 & 1.74 & 33.1 & $32 (2.3)$& $17  (2.3)$& $17.9      $\\
\hline
\end{tabular}
\vspace*{0.5cm}
\caption[]{Some properties as a function of the attempt frequency $f$:
$\xsamp \equiv \tsamp/\tstar(f)$ with $\tsamp=10^5$ and $\tstar(f)=16/f$,
excess energy per bead $e$,
average normal pressure $P$,
affine shear elasticity $\muA$,
contributions $\muFtild$ and $\muFstar$ to the shear-stress fluctuation $\muF=\muFtild-\muFstar$
and $\GF = \muA-\muF$. All data are averaged over $m=100$ configurations. 
Error bars (for values $>0.1$) are indicated for the last three columns.
\label{tab}}
\end{table}

In addition to the MC moves changing the connectivity matrix of the network 
standard velocity-Verlet molecular dynamics (MD) \cite{AllenTildesleyBook} 
is used to move the beads through the phase space.
The temperature $T=1$ is imposed using a Langevin thermostat of friction constant $\zeta=1$.
This allows to suppress long-range hydrodynamic modes otherwise relevant for two-dimensional systems.
A velocity-Verlet time step $\dtMD = 10^{-2}$ is used. Every time step $\dtMD$ a certain 
number of springs corresponding to the frequency $f$ is considered for an MC hopping move.
We start by equilibrating $m=100$ independent configurations at $f=1$. 
The frequency is then decreased with steps $f=1, 0.3, 0.1, 0.03, 0.01, \ldots , 10^{-7}$ and finally $f=0$.
At each step the configurations are tempered over a time interval $\ttemp=10^4$ and then sampled over $\ttraj=10^5$. 
Due to detailed balance changing $f$ does not change the standard static properties, 
such as described by the pair correlation functions $g(r)$ and $S(q)$ (Fig.~\ref{fig_pairV}) 
or the energy per bead $e$ or the normal pressure $P$ shown in Table~\ref{tab}.
As we have checked, one could have also considered a much more rapid quench 
without changing these static properties.
As seen from Fig.~\ref{fig_algo_snap}, we obtain homogeneous and isotropic elastic networks well 
above the percolation threshold \cite{DegennesBook,StaufferBook}. 
This is consistent with the large values $\GF \approx 18$ for small $f$ 
in Table~\ref{tab}.

\section{Computational results}
\label{sec_simu}

\subsection{Static and quasi-static properties}
\label{simu_static}

\begin{figure}[t]
\centerline{\resizebox{1.0\columnwidth}{!}{\includegraphics*{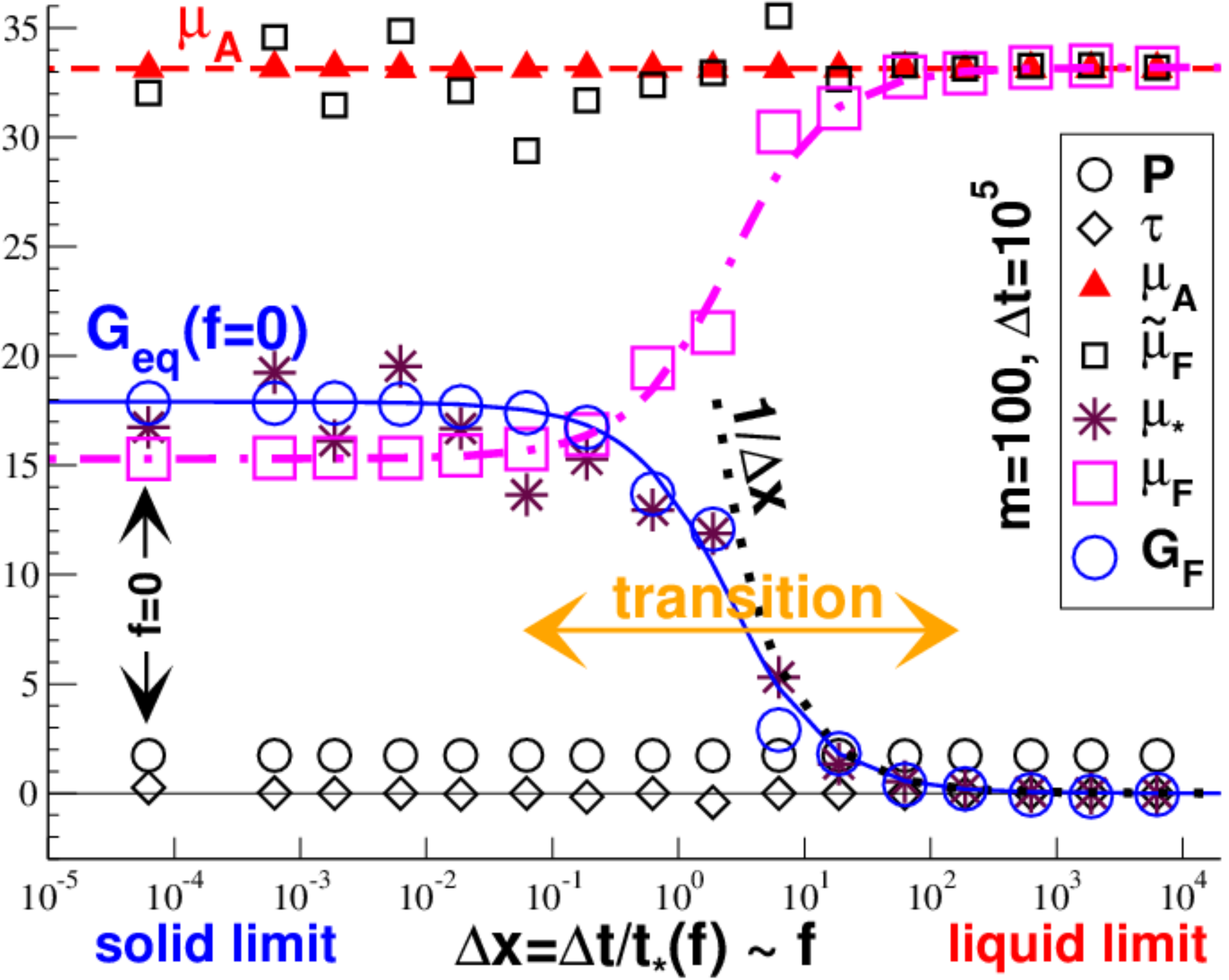}}}
\caption{Various ``static" and ``quasi-static" properties {\em vs.} 
$\xsamp(f) \equiv \tsamp/\tstar(f)$ with $\tsamp=10^5$ and $\tstar(f)=16/f$.
The data indicated for the smallest $\xsamp$ correspond to $f=0$.
The prediction Eq.~(\ref{eq_GFfDebye}) is indicated by the solid and the dash-dotted lines.
Note that $\muFtild \approx \muA$ and $\muFstar \approx \GF$ for all $\xsamp$.
\label{fig_static_f}
}
\end{figure}

We begin the description of our transient networks by discussing the static and 
quasi-static properties presented in Fig.~\ref{fig_static_f}.
For every attempt frequency $f$ we sample $m=100$ configurations over a fixed sampling time 
$\tsamp=\ttraj=10^5$. 
For each trajectory we store every $\dtMD=0.01$ instantaneous
properties such as the normal pressure $\Phat$, the shear stress $\tauhat$ or the affine shear 
elasticity $\muAhat$ as defined in Appendix~\ref{theo_muAhat}. Using these instantaneous values $\ahat$ 
we then sample the time averages
$\overline{\ahat}$ and $\overline{\ahat^2}$ over the $\tsamp/\dtMD$ entries for each configuration.
Using these time averages we obtain for each configuration an observable $\Ocalhat$
and compute its first moment $\Ocal = \langle \Ocalhat \rangle$ over the $m$ configurations.
(The second moment $\langle \Ocalhat^2 \rangle$ will be considered in Sec.~\ref{simu_mbound}.)
The following properties 
\begin{eqnarray}
\Ocalhat = \overline{\Phat} & \Rightarrow & \Ocal = P  \label{eq_P_av} \\
\Ocalhat = \overline{\tauhat} & \Rightarrow & \Ocal = \tau  \label{eq_tau_av} \\
\Ocalhat = \overline{\muAhat} & \Rightarrow & \Ocal = \muA  \label{eq_muA_av} \\
\Ocalhat = \beta V \overline{\tauhat^2}   & \Rightarrow & \Ocal = \muFtild \label{eq_muFtild_av} \\
\Ocalhat = \beta V \overline{\tauhat}^2   & \Rightarrow & \Ocal = \muFstar \label{eq_muFstar_av} \\
\Ocalhat = \beta V (\overline{ \tauhat^2} - \overline{\tauhat}^2)   & \Rightarrow & \Ocal = \muF
= \muFtild - \muFstar \label{eq_muF_av} \\
\Ocalhat = \overline{\muAhat} - \beta V (\overline{ \tauhat^2} - \overline{\tauhat}^2) & \Rightarrow & \Ocal = \GF = \muA - \muF \label{eq_GF_av}
\end{eqnarray}
are presented in Fig.~\ref{fig_static_f} using log-linear coordinates.
The vertical axis has the dimension energy per volume. The horizontal axis has been made dimensionless 
using  $\xsamp \equiv \tsamp/\tstar(f)$ with $\tstar(f)=16/f$ as shown below in Sec.~\ref{simu_MSD}.
(See Appendix~\ref{simu_dt} for the scaling with sampling time $\tsamp$ at fixed $f$.)
In the ``solid limit" ($\xsamp \ll 1$) only few spring recombinations can occur and the networks 
thus behave as solid bodies, while in the ``liquid limit" ($\xsamp \gg 1$) the
particles may freely change their neighbors.

As may be seen from Table~\ref{tab} or Fig.~\ref{fig_static_f}, the pressure $P$, 
the shear stress $\tau$, the affine shear elasticity $\muA$ and the contribution $\muFtild$ 
to the shear-stress fluctuation $\muF$ do not depend on $\xsamp$, i.e. the same values $P \approx 1.7$,
$\tau \approx 0$ and $\muA \approx \muFtild \approx 33.2$ have been obtained for all $f$. 
The expectation values of these truly ``static" properties cannot depend on 
$\tsamp$ or on $f$ since time and ensemble averages do ``commute" \cite{WXB16}, i.e. can be exchanged as
\begin{equation}
\la \overline{\ahat} \ra = \overline{\la \ahat \ra}
\mbox{ with } \ahat = \Phat, \tauhat, \muAhat \mbox{ or } \beta V \tauhat^2,
\label{eq_commute} 
\end{equation}
and since the thermodynamic ensemble average $\langle \ldots \rangle$ does not depend on $\tsamp$ or $f$.
Although $P$, $\tau$, $\muA$ and $\muFtild$ are all $\xsamp$-independent,
this does not imply that they have the same statistics. The ``simple averages" $P$, $\tau$ and $\muA$ 
have been obtained with a high precision while the ``fluctuation" $\muFtild$ is rather noisy 
\cite{AllenTildesleyBook,WXB16}.

A qualitatively different behavior is observed for the observables $\muFstar(\xsamp)$, $\muF(\xsamp)$ and $\GF(\xsamp)$
also represented in Fig.~\ref{fig_static_f}. Please note that Eq.~(\ref{eq_commute})
does not hold for these properties as may be seen for 
$\muFstar(\xsamp) = \langle \overline{s}^2 \rangle \ge 0$ with $s = \sqrt{\beta V} \tauhat$.
Obviously, this differs from $\overline{\langle s \rangle^2} \sim \tau^2$ which vanishes due 
to symmetry for all $\xsamp$ for a sufficiently large ensemble.
Ergodicity implies $\overline{s} \to \langle s \rangle$ for large $\xsamp$ and all 
$\xsamp$-effects become thus irrelevant. As seen from Fig.~\ref{fig_static_f}, 
this implies $\muFstar(\xsamp) \to \beta V\tau^2 = 0$ for $\xsamp \gg 1$. 
Similarly, one observes $\GF(\xsamp) = \muA - \muF(\xsamp) \to 0$ and thus $\muF(\xsamp) \to \muA$
as expected for liquids \cite{WXP13,WXB15}. 
The quasi-static properties become also constant for $\xsamp \ll 1$ where  
$\GF(\xsamp) \to \Geq(f=0) \approx 18$ and
$\muF(\xsamp) \to \muF(f=0) \approx 15.$
Interestingly, the transition between both limits around $\xsamp \approx 1$ is rather broad
corresponding to several orders of magnitude. 
Our data are nicely fitted over the full range of $\xsamp$ by the expected behavior Eq.~(\ref{eq_GFfDebye}) 
for a Maxwell fluid as indicated by the thin solid line for $\GF(\xsamp) \approx \muFstar(\xsamp)$ and 
by the dash-dotted line for $\muF(\xsamp) = \muA - \GF(\xsamp)$.
We remind that $\fDebye(x) \to 1$ for $x \to 0$ and $\fDebye(x) \to 2/x$ for $x \gg 1$.
This implies that  $\GF(\xsamp)$ decays as $2 \Gstar/\xsamp$ in the liquid limit 
as shown by the dotted line. 

Let us finally consider the scaling of the two contributions $\muFtild$
and $\muFstar(\xsamp)$ to the shear-stress fluctuation $\muF(\xsamp)$.
As seen from Fig.~\ref{fig_static_f}, we have $\muFtild \approx \muA$
in agreement with Eq.~(\ref{eq_condition}) and in addition
\begin{equation}
\muFstar(\xsamp) \approx \GF(\xsamp) \label{eq_GFmuFstar}
\end{equation}
for all $\xsamp$. 
As already stressed, the expectation value of $\muFtild(\xsamp)$ does not depend on $\xsamp$. 
This must especially hold for large $\xsamp$ where the average shear stress 
$\overline{s} \equiv \sqrt{\beta V} \ \overline{\tauhat}$ must vanish for each configuration 
and, hence, $\muFstar(\xsamp) = \langle \overline{s}^2 \rangle \approx 0$.
Since the stress-fluctuation estimate $\GF$ for the shear modulus, Eq.~(\ref{eq_GF}), 
must also vanish in the liquid limit, this implies $0 \approx \muA -\muFtild$ for $\xsamp \gg 1$.
Since $\muFtild$ does not depend on $\xsamp$, this demonstrates Eq.~(\ref{eq_condition})
and using Eq.~(\ref{eq_GF}) this implies in turn Eq.~(\ref{eq_GFmuFstar}). 
Please note that Eq.~(\ref{eq_condition}) and Eq.~(\ref{eq_GFmuFstar}) do not hold for an arbitrary 
elastic body as shown, e.g., in Ref.~\cite{WXB16}. In fact they do not necessarily hold even for {\em one} 
configuration of our ensemble if $\xsamp \ll 1$. As we shall see in Sec.~\ref{simu_mbound}, they only apply 
for $\xsamp \gg 1$ or for an average over a large number $m$ of configurations for $\xsamp \ll 1$.

\begin{figure}[t]
\centerline{\resizebox{0.9\columnwidth}{!}{\includegraphics*{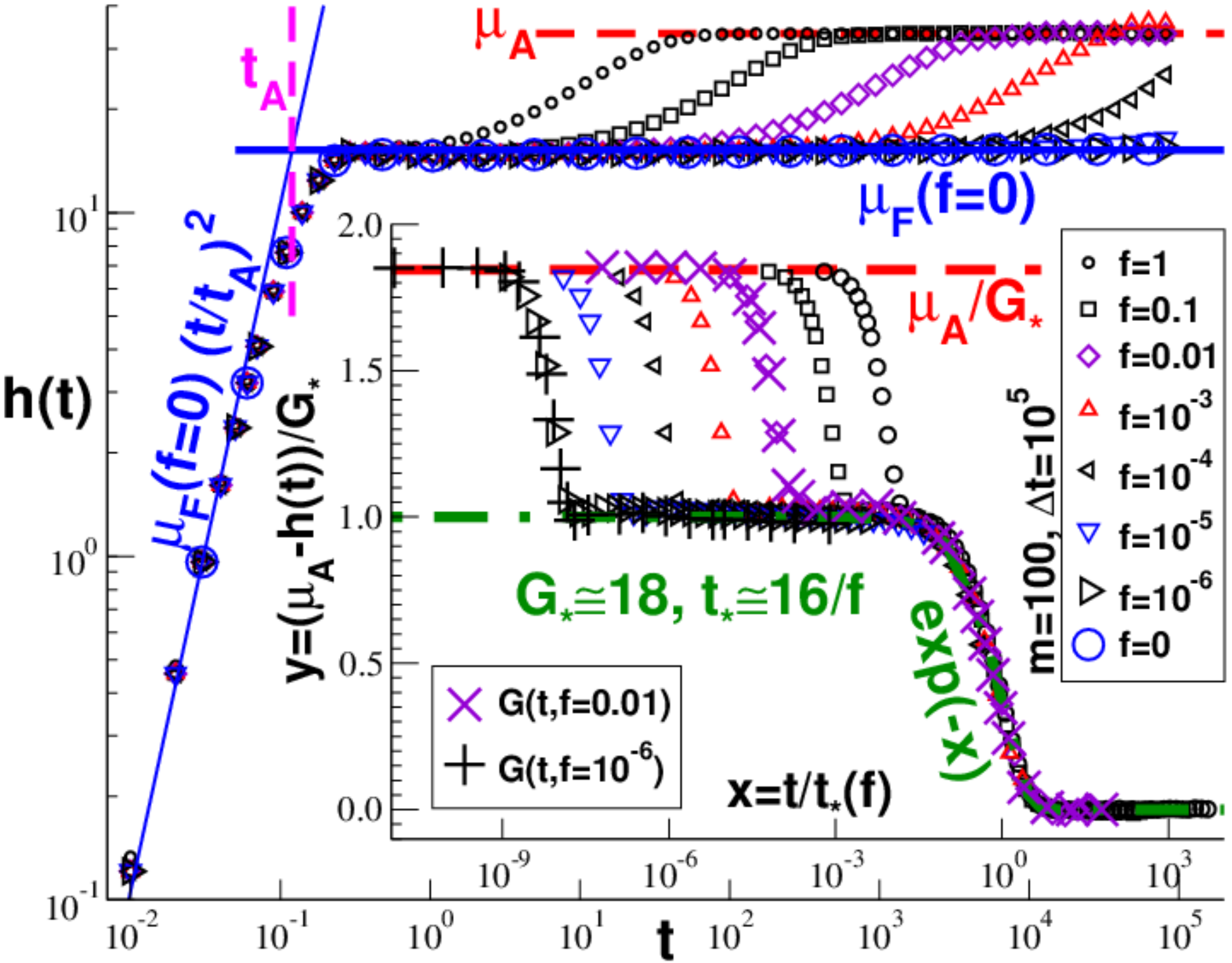}}}
\caption{(Color online)
Shear-stress MSD $h(t)$ for different frequencies $f$.
Main panel:
The MSD increases as $h(t) \sim t^2$ for small times $t \ll \tauA$ (thin solid line),
shows an intermediate plateau with $h(t) \approx \muF(f=0)$ for $\tauA \ll t \ll \tstar(f)$ 
(bold solid line) and approaches $\muA$ (dashed line) for even larger times $t \gg \tstar(f)$.
Inset:
Comparison of $y(x)=(\muA-h(t))/\Gstar$ with $x=t/\tstar(f)$, $\tstar(f)=16/f$ and 
$\Gstar=\Geq(f=0)$ with the shear-stress response modulus $G(t)/\Gstar$ for $f=0.01$ and $f=10^{-6}$
obtained from the shear-stress increment $\la \delta \tauhat(t) \ra$ 
after applying a step-strain increment $\delta \gamma = 0.01$ at $t=0$. 
\label{fig_MSD_f}
}
\end{figure}

\subsection{Shear-stress mean-square displacement}
\label{simu_MSD}

The shear-stress MSD $h(t)$ is presented in Fig.~\ref{fig_MSD_f} for a broad range of 
attempt frequencies $f$. The data have been computed using 
\begin{equation}
h(t) \equiv \frac{\beta V}{2} 
\la \overline{\left(\tauhat(t+t_0) - \tauhat(t_0) \right)^2} \ra
\label{eq_MSD_def}
\end{equation}
where the horizontal bar stands for the gliding average over $t_0$ \cite{AllenTildesleyBook}
for each configuration using a fixed time window $\tsamp=10^5$ and
$\la \ldots \ra$ for the ensemble average over $m=100$ configurations.
Time and ensemble averages commute, Eq.~(\ref{eq_commute}), i.e. the expectation value of the MSD 
does not depend explicitly on the sampling time as emphasized in Ref.~\cite{WXB16}.
Let us focus first on the main panel of Fig.~\ref{fig_MSD_f} where the unscaled $h(t)$ 
is presented using double-logarithmic coordinates. Three dynamical regimes can be distinguished
corresponding to the time windows {\em (i)} $t \ll \tauA$, {\em (ii)} $\tauA \ll t \ll \tstar(f)$
and {\em (iii)} $\tstar(f) \ll t$. The MSD does not depend on the attempt frequency $f$
in the first two regimes, i.e. the reorganization of the spring network is still irrelevant.
The two indicated solid lines form a lower envelope for $h(t)$ for $f \to 0$.
The MSD increases as $h(t) \sim t^2$ in the first regime \cite{foot_powertwo}
and shows an intermediate plateau with $h(t) \approx \muF(f=0)$ in the second. 
Following Refs.~\cite{WKB15,WXB16} the value of the crossover time $\tauA \approx 0.12$ 
is fixed by matching the asymptotics as indicated by the vertical dash-dotted line.
The second regime is consistent with the equilibrium modulus of the quenched network 
$\Geq(f=0) = \GF(f=0) \approx \muA - h(t) \approx 18$ for $\tauA \ll t \ll \tstar(f)$.
The spring recombinations become relevant for times of order $\tstar(f)$.
Depending on $f$ the MSD $h(t)$ increases now further approaching from
below the long time limit $h(t) \to \muF(f>0) = \muA$ and
the $f$-dependence thus drops out again. 

We have yet to verify the scaling of the network relaxation time $\tstar(f)$
which characterizes the crossover from the second to the third regime.
This is done in the inset of Fig.~\ref{fig_MSD_f} where $h(t)$ is 
replotted using a half-logarithmic representation.
The axes are made dimensionless by plotting $y(x)=(\muA-h(t))/\Gstar$ as a function
of the reduced time $x \equiv t/\tstar(f)$ where we set $\Gstar \equiv \Geq(f=0)$
for the intermediate plateau modulus and $\tstar(f) \equiv 16/f$ for the network relaxation time.
This rescaling leads to a perfect collapse of the data for $x \gg \xAf \equiv \tauA/\tstar(f)$,
especially for the $f$-dependent regime seen in the main panel.
Moreover, the reduced MSD is seen to decay exponentially as $y(x)=\exp(-x)$ for $x \gg \xAf$ 
(dash-dotted line). 
The prefactor $16$ for $\tstar(f)$ has been introduced for convenience.
For not too small attempt frequencies $f \ge 10^{-4}$, the exponential decay and 
the scaling of the relaxation time $\tstar(f)$ may also be checked by plotting 
the unscaled $\muA-h(t)$ {\em vs.} $t$ using a linear-logarithmic representation (not shown).

Due to the uncorrelated recombinations of the springs a Maxwell fluid relaxation 
is expected for our simple model. The observed exponential decay, Eq.~(\ref{eq_Maxwell}), 
thus confirms Eq.~(\ref{eq_key}). This is also demonstrated 
by the comparison with the directly computed relaxation moduli for the two 
attempt frequencies $f=0.01$ and $f=10^{-6}$ corresponding, respectively, to the liquid limit 
($\xsamp=62.5 \gg 1$) and the solid limit ($\xsamp=0.00625 \ll 1$). 
As in our recent studies on permanent elastic networks \cite{WXB15,WXBB15,WKB15,WXB16}
the relaxation modulus has been computed from the shear-stress increment 
$\la \delta \tauhat(t) \ra$ with $\delta \tauhat(t) \equiv \tauhat(t) - \tauhat(0^{-})$ 
measured after a step-strain $\delta \gamma = 0.01$ has been applied at $t=0$. 
This was done by applying a canonical-affine shear transformation (Appendix~\ref{theo_affine})
and by averaging over $m=100$ independent configurations. 
The perfect data collapse for all times confirms Eq.~(\ref{eq_key}).

\begin{figure}[t]
\centerline{\resizebox{0.9\columnwidth}{!}{\includegraphics*{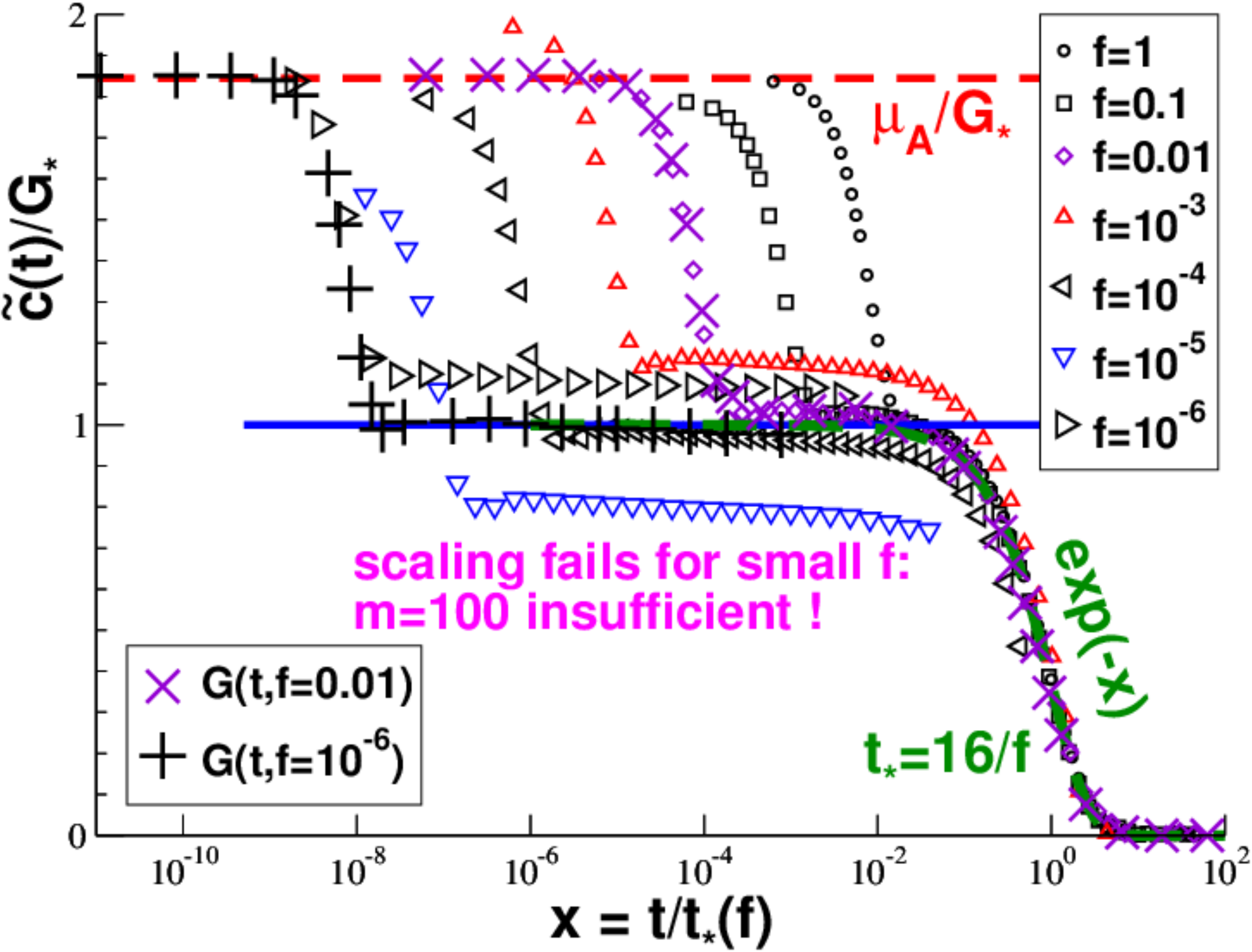}}}
\caption{(Color online)
Rescaled shear-stress ACF $\Cttild/\Gstar$ {\em vs.} dimensionless time
$x = t/\tstar(f)$ for a broad range of $f$. Also indicated are the similarly
rescaled relaxation moduli $G(t)$ obtained for $f=0.01$ and $f=10^{-6}$
by applying a step strain $\delta \gamma=0.01$. The scaling clearly
fails for small $f$ (small $\xsamp$). 
\label{fig_ACF_f}
}
\end{figure}

\subsection{Shear-stress auto-correlation function}
\label{simu_ACF}
Instead of using the MSD $h(t)$ the response modulus is generally estimated 
in computational studies using the shear-stress ACF 
$\Cttild \equiv \beta V \langle \overline{\tauhat(t+t_0) \tauhat(t_0)} \rangle$
presented in Fig.~\ref{fig_ACF_f}. Time and ensemble averages do again commute
and the expectation value does thus not depend on $f$ or $\tsamp$.
As suggested in Ref.~\cite{WXB16}, one can instead of using the MSD $h(t)$ and 
Eq.~(\ref{eq_key}) equivalently determine the relaxation modulus using 
\begin{equation}
G(t) = \muA - \muFtild + \Ctild(t).
\label{eq_GtACFtild}
\end{equation}
This is justified under the condition that the {\em measured} values for $\muA$ 
and $\muFtild$ for each $f$ are taken. Due to the exact identity \cite{DoiEdwardsBook} 
\begin{equation}
h(t) = \Ctild(0) - \Ctild(t) = \muFtild - \Ctild(t) 
\label{eq_htCt}
\end{equation}
this yields precisely the same results (not shown) as already presented in the inset of Fig.~\ref{fig_MSD_f}.
Please note that for a general solid body, $\muA-\muFtild$ may be very different from zero and 
cannot be neglected in general \cite{WXB16}. Albeit the expectation value of this difference 
(obtained for asymptotically large $m$ or $\xsamp$) does vanish for any liquid (Fig.~\ref{fig_static_f}),
the difference found for $m=100$ configurations is apparently not small enough.
This explains the bad scaling for small $f$ shown in Fig.~\ref{fig_ACF_f} where 
$\Cttild/\Gstar$ is traced as a function of $x=t/\tstar$ as in the inset of Fig.~\ref{fig_MSD_f}.
The approximation Eq.~(\ref{eq_keyapprox}) thus does not have the same status as the fundamental 
relation Eq.~(\ref{eq_key}). 

\begin{figure}[t]
\centerline{\resizebox{1.0\columnwidth}{!}{\includegraphics*{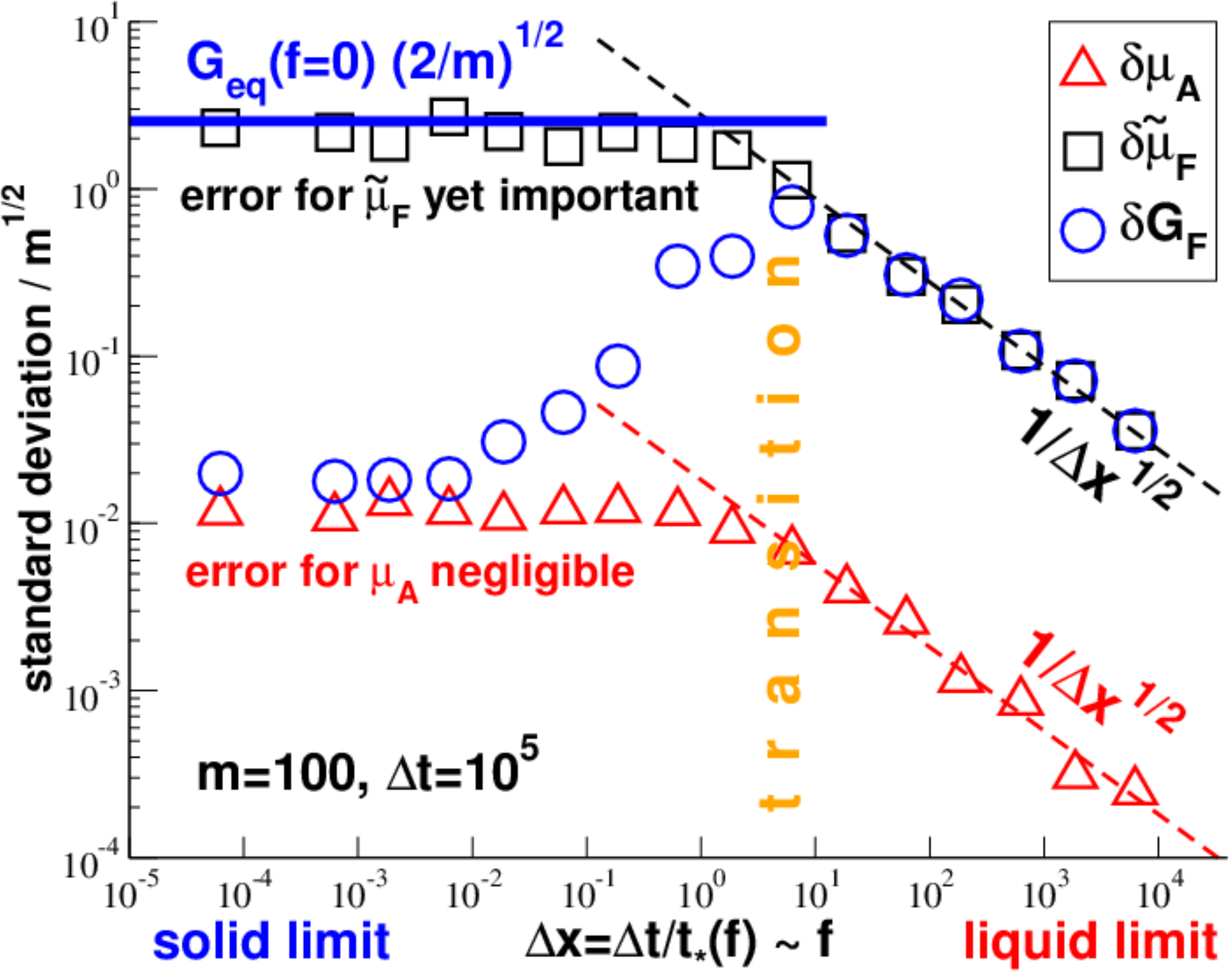}}}
\caption{(Color online)
Error bars $\delta o/\sqrt{m}$ for $o = \muA$, $\muFtild$ and $\GF=\muA-\muF$ 
as a function of $\xsamp = \tsamp/\tstar(f)$.
The error bars for $\muA$ are several orders of magnitude smaller than those for $\muFtild$.
The deviations from $\muA-\muFtild=0$ observed in Fig.~\ref{fig_ACF_f} 
for small $f$ are thus due to the fluctuations of $\muFtild$.
\label{fig_noise_f}
}
\end{figure}

\subsection{Minimal number of configurations required} 
\label{simu_mbound}

Using the $m$ independent configurations for each $f$ we have computed the standard deviations 
$\delta \Ocal \equiv (\langle \Ocalhat^2 \rangle - \langle \Ocalhat \rangle^2)^{1/2}$
and error bars $\delta \Ocal/\sqrt{m}$ associated with the average properties 
$\langle \Ocalhat \rangle$ discussed above.
Let us first summarize the standard deviations $\delta \muA$, $\delta \muFtild$ and $\delta \GF$ 
associated to $\muA$, $\muFtild$ and $\GF = \muA-\muF$.
The corresponding error bars are traced in Fig.~\ref{fig_noise_f}.
As one expects assuming an increasing number $\propto \xsamp$ of independent networks 
probed by each configuration, all properties decay as $1/\sqrt{\xsamp}$ (dashed lines) 
in the liquid limit ($\xsamp \gg 1$). Note that $\delta \muA$ and $\delta \muFtild$ 
become constant for $\xsamp \ll 1$ where each configuration only probes
one network topology. As indicated by the bold horizontal line \cite{foot_notdone}, 
\begin{equation}
\delta \muFtild \approx \sqrt{2} \Geq(f=0) \mbox{ for } \xsamp \ll 1.
\label{eq_dmuFtild_bound}
\end{equation}
Interestingly, $\delta \GF$ reveals a qualitatively different non-monotonous behavior with 
a clear maximum at the transition at $\xsamp \approx 1$ between the liquid and the solid limit.
While $\delta \GF \approx \delta \muFtild$ for $\xsamp \gg 1$, $\delta \GF$ becomes several 
orders of magnitude smaller than $\delta \muFtild$ for $\xsamp \ll 1$ and even becomes similar 
to $\delta \muA$ for very small $\xsamp$.
More details on the fluctuations of static properties (especially on their scaling with
system size) will be given elsewhere. 

\begin{figure}
\centerline{\resizebox{1.0\columnwidth}{!}{\includegraphics*{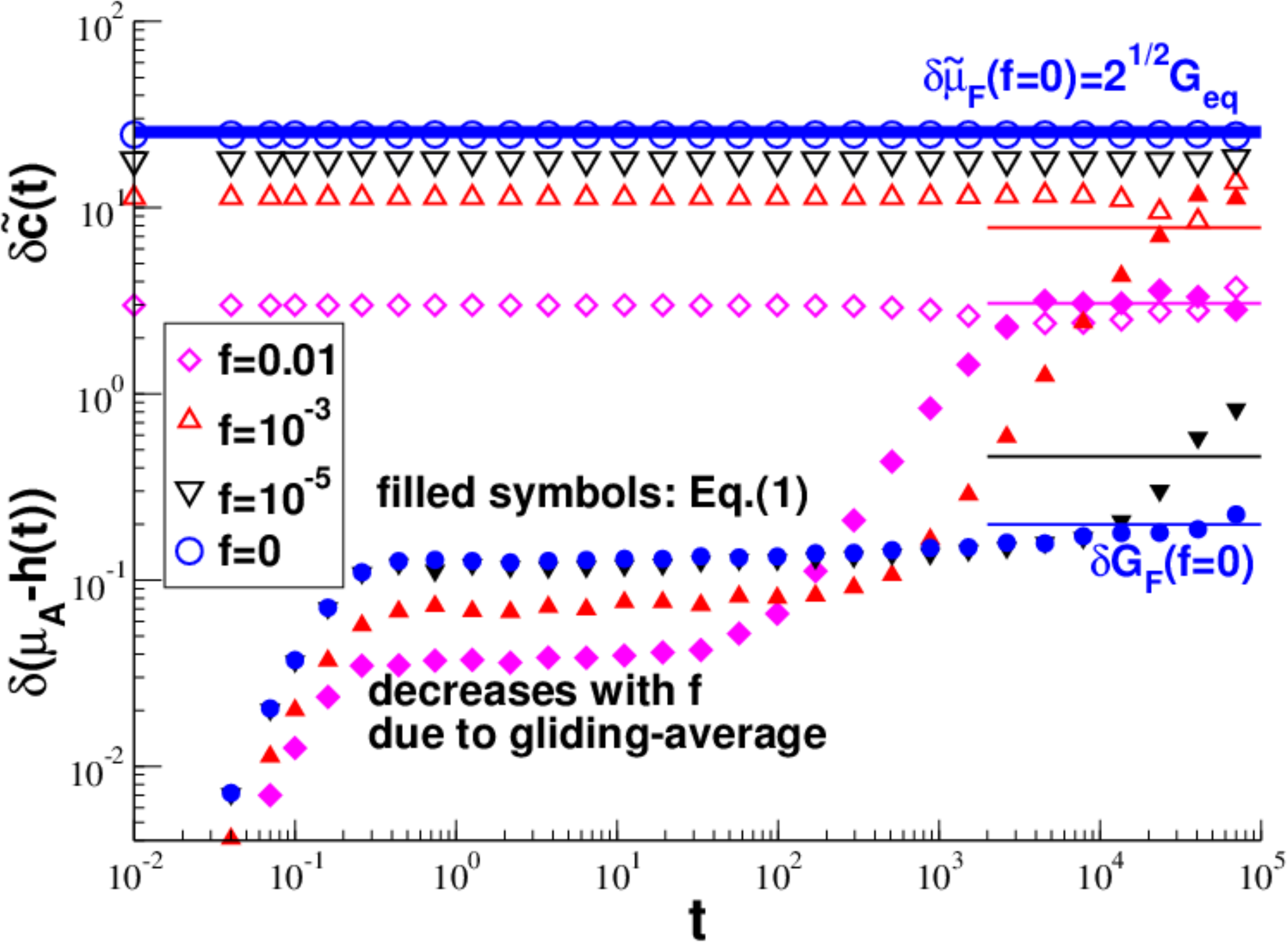}}}
\caption{(Color online)
Standard deviations $\delta (\muA-h(t))$ (filled symbols) and $\delta \Cttild$ (open symbols)
for several attempt frequencies $f$ as indicated. 
While $\delta \Cttild$ becomes similar
to this bound for $\xsamp \ll 1$, $\delta (\muA-h(t))$ is orders of 
magnitude smaller in the same limit. 
The thin horizontal lines indicate $\delta \GF(f)$ for $f=0$ (bottom), 
$f=10^{-5}$, $f=0.01$ and $10^{-3}$ (top). $\delta (\muA-h(t))$ is seen 
to approach this limit for $t \to \tsamp$.
}
\label{fig_errGt}
\end{figure}

Figure~\ref{fig_errGt} presents the standard deviations $\delta (\muA-h(t))$ and $\delta \Cttild$ 
associated with Eq.~(\ref{eq_key}) and Eq.~(\ref{eq_keyapprox}).
$\delta \Cttild$ is apparently time-independent. 
One verifies that
\begin{equation}
\delta \Cttild \approx \delta (\muA- \muFtild) \approx \delta \muFtild,
\label{eq_dCttild}
\end{equation}
i.e. the noise is set by the fluctuations of the neglected term $\muA-\muFtild$.
(As known from Fig.~\ref{fig_noise_f}, $\delta \muA$ is negligible.)
The limit Eq.~(\ref{eq_dmuFtild_bound}) for $\delta \muFtild$ is thus
also an upper bound for $\delta \Cttild$ (bold horizontal line).
The time-dependence of $\delta (\muA-h(t))$ is more intricate (filled symbols).
One (slightly trivial) reason for this is that gliding averages are used, 
Eq.~(\ref{eq_MSD_def}), which reduce more efficiently the fluctuations for short times
and higher frequencies (where more statistically independent networks are probed).
We thus observe that $\delta (\muA-h(t)) \approx \delta h(t)$ increases monotonously with time.
It becomes similar to $\delta \GF(f)$ for $t \to \tsamp$ as indicated
by thin horizontal lines.
We emphasize that $\delta (\muA-h(t))$ is several orders of magnitude smaller than
$\delta \Cttild$ for most times $t$ and attempt frequencies $f$. Both fluctuations 
become similar only for large times $t \approx \tsamp$ in the liquid limit above 
$\xsamp \approx 1$.

The goal is now to characterize roughly the lower bound $\mbound$ of configurations required 
for a given $\xsamp$ for both methods Eq.~(\ref{eq_key}) and Eq.~(\ref{eq_keyapprox}). 
Let us suppose that the relaxation modulus $G(t)$ is needed with a fixed precision $\Gpr$, 
say $\Gpr=1$.
As explained above, the problem with Eq.~(\ref{eq_keyapprox}) is that $\muFtild$ is a strongly 
fluctuating quantity. Using Eq.~(\ref{eq_dCttild})
this leads to the criterion
\begin{equation}
m \gg \mbound = (\delta \muFtild/\Gpr)^2 \ \mbox{ for Eq.~(\ref{eq_keyapprox}).}
\label{eq_mcriterion}
\end{equation}
According to the upper limit Eq.~(\ref{eq_dmuFtild_bound}) this corresponds to
a minimal number of $\mbound = 2 (\Geq/\Gpr)^2 \approx 650$ configurations in the
solid limit which exceeds by nearly an order of magnitude the number of
configurations we have been able to simulate. This is consistent with the bad scaling 
found in this limit in Fig.~\ref{fig_ACF_f}.
As shown in Fig.~\ref{fig_errGt} $\delta (\muA-h(t))$ is monotonously increasing 
with time approaching $\delta \GF(f)$ from below. Replacing the detailed
time dependence by this upper limit yields the simple, albeit rather conservative criterion
\begin{equation}
m \gg \mbound = (\delta \GF/\Gpr)^2 \ \mbox{ for Eq.~(\ref{eq_key}).}
\label{eq_GFmcriterion}
\end{equation}
Both criteria are identical in the liquid limit where $\delta \muFtild \approx \delta \GF$. 
However, Eq.~(\ref{eq_GFmcriterion}) corresponds to a pronounced
maximum at $\xsamp \approx 1$ and decreases then by several
orders of magnitude if we enter further into the solid limit.
Note that the bound $\mbound$ implied by Eq.~(\ref{eq_GFmcriterion}) remains everywhere below $m=100$. 
This is consistent with the excellent statistics observed in Fig.~\ref{fig_MSD_f} for all $f$.

\section{Conclusion}
\label{sec_conc}

\subsection{Summary}
\label{conc_summary}

The present study had two main goals.
One was to introduce a simple generic model for self-assembled elastic networks 
(Sec.~\ref{sec_algo}) and to characterize it numerically (Sec.~\ref{sec_simu}).
In this model repulsive beads are reversibly bridged by ideal springs 
which recombine locally with an MC attempt frequency $f$ (Fig.~\ref{fig_sketch}).
By construction our transient networks are Maxwell fluids \cite{RubinsteinBook} 
with a longest relaxation time $\tstar(f) \sim 1/f$ and an intermediate plateau modulus
$\Gstar$ given by the equilibrium shear modulus $\Geq$ for quenched network topologies ($f=0)$.
By varying the dimensionless attempt frequency $\xsamp = \tsamp/\tstar$
one may thus scan continuously between the liquid limit ($\xsamp \gg 1$) 
and the solid limit ($\xsamp \ll 1$). 
This was done by varying the attempt frequency $f$ (Figs.~\ref{fig_pairV}-\ref{fig_noise_f}) 
and, more briefly, by changing the sampling time $\tsamp$ (Figs.~\ref{fig_GF_dt} 
and \ref{fig_Gt_fmid}).
Due to detailed balance all static properties related to particle pair correlations 
(Fig.~\ref{fig_pairV}) are kept constant. This is different for the quasi-static 
properties $\muFstar(\xsamp)$, $\muF(\xsamp)$ and $\GF(\xsamp)$ due to the finite 
time needed for stress fluctuations to explore the phase space (Fig.~\ref{fig_static_f}). 
The $\xsamp$-dependence of these properties are perfectly described by the prediction 
Eq.~(\ref{eq_GFfDebye}) made for Maxwell fluids.

The second goal of this work was to use this deliberately simple model to verify 
(Fig.~\ref{fig_MSD_f}) the simple-average relation Eq.~(\ref{eq_key}) recently proposed 
for the computational determination of the shear-stress relaxation modulus $G(t)$ \cite{WXB16}. 
An alternative derivation of Eq.~(\ref{eq_key}) was given (Appendix~\ref{theo_Gt}) 
which does not rely on the steepest-descend assumption implicit to the Lebowitz-Percus-Verlet 
transformation between conjugated ensembles \cite{AllenTildesleyBook,Lebowitz67} used in 
our previous work \cite{WXB15,WXBB15,WKB15,WXB16}. 
The formula Eq.~(\ref{eq_key}) has been compared (Fig.~\ref{fig_ACF_f}) with the generally 
assumed Eq.~(\ref{eq_keyapprox}) using only the shear-stress autocorrelation function $\Cttild$.
While from the theoretical point of view the latter relation is applicable for liquids since 
Eq.~(\ref{eq_condition}) holds on average (Fig.~\ref{fig_static_f}), it imposes severe restrictions
on computational studies due to the large fluctuations of $\muFtild$ (Fig.~\ref{fig_noise_f}). 
This implies that at least $\mbound \approx 2 (\Geq/\Gpr)^2$ independent configurations are needed 
for $\xsamp \ll 1$.
At contrast to this Eq.~(\ref{eq_key}) provides an approximation-free alternative with a much 
better statistics in the solid limit (Fig.~\ref{fig_errGt}).

\subsection{Outlook}
\label{conc_outlook}
The present study has focused on the variation of the attempt frequency $f$ while keeping fixed 
other parameters such as the volume $V$, the bead density $\rho$, 
the spring density $\rhosp$ or the temperature $T$. 
It should be particularly rewarding to systematically investigate system size effects.
While most properties discussed here, such as $\muA$, $\muFtild$, $h(t)$ or $\Cttild$, 
are defined such that their expectation values, i.e. their first moments 
over the ensemble of independent configurations, should not depend explicitly on $V$,
this is less obvious for their respective standard deviations. 
As stated by the criterion Eq.~(\ref{eq_mcriterion}), we expect a strong lack of self-averaging 
\cite{AllenTildesleyBook,LandauBinderBook} for $\delta \muFtild$, i.e. the approximation 
Eq.~(\ref{eq_keyapprox}) should not improve with increasing system size, while strong
self-averaging is expected for Eq.~(\ref{eq_key}) in the low-$\xsamp$ limit.
As already stated in the Introduction, our transient networks are rheologically similar to the 
Maxwell fluids formed by patchy colloids \cite{Leibler13,Kob13} or by so-called ``vitrimers" 
\cite{Leibler11}.
Interestingly, these physical gels can be reworked (just as silica glasses) to any shape 
by tuning gradually the system temperature $T$ which is the central experimental control parameter.
Since our model potentials are rather stiff (Fig.~\ref{fig_algo_pot}), changing slightly $T$ will 
not alter much the local static structure, i.e., $\muA$ and $\Geq$ should remain essentially constant. 
However, by assuming the MC attempt frequency $f$ of our transient networks to be thermally 
activated, i.e. $f(T) \sim \exp(-B/T)$ as for patchy colloids \cite{Leibler13}, this should 
imply a strong Arrhenius behavior for the Maxwell relaxation time $\tstar(f)$ and the shear viscocity
\begin{equation}
\eta \approx \Gstar \ \tstar(f) \sim 1/f \sim \exp(B/T).
\label{eq_visco}
\end{equation}
Our networks should thus behave as ``strong glasses" \cite{HansenBook}. 

\vspace*{0.2cm} 
\begin{acknowledgments}
H.X. and I.K. thank the IRTG Soft Matter for financial support.
We are indebted to C.~Ligoure (Montpellier) and J.~Farago (Strasbourg) for helpful discussions.
\end{acknowledgments}

\appendix
\section{Canonical-affine shear transformation}
\label{theo_affine}

Let us apply an infinitesimal shear strain increment $\gamma \to \gamma + \delta \gamma$ to a 
periodic simulation box at constant box volume $V$ at a reference shear strain $\gamma$.
(For simplicity all particles are in the principal box \cite{AllenTildesleyBook}.)
The positions $\rvec_i$ and the velocities $\vvec_i$ \cite{foot_monodisp} of all particles $i$ 
are assumed to follow the ``macroscopic" constraint in a both {\em affine} \cite{WTBL02,TWLB02}
and {\em canonical} \cite{Goldstein} manner according to 
\begin{equation}
\rix \to \rix + \delta \gamma \ \riy \ , \ \vix \to \vix - \delta \gamma \ \viy 
\label{eq_cantrans}
\end{equation}
with $|\delta \gamma| \ll 1$. All other coordinates and velocities remain 
unchanged by the transformation as well as the network of springs connecting the particles.
The negative sign for the velocities assures that the transform is ``canonical" \cite{foot_monodisp} 
and that, hence, Liouville's theorem is obeyed \cite{Goldstein,WKB15}. The transformation 
Eq.~(\ref{eq_cantrans}) is used in Sec.~\ref{simu_MSD} to test our key relation Eq.~(\ref{eq_key}).

\section{Shear stress and affine shear elasticity}
\label{theo_muAhat}

Let $\Hhat(\state,\gamma)$ denote the system Hamiltonian of a given state $\state$ 
at an imposed shear strain $\gamma$ of the simulation box.
The state $\state$ of the system specifies the positions and velocities of the
particles and the connectivity matrix of the ideal springs connecting them. The two configurations
shown in panel (c) of Fig.~\ref{fig_sketch} thus correspond to two different states.
The instantaneous shear stress $\tauhat$ and the instantaneous affine shear elasticity $\muAhat$ 
are defined by functional derivatives of the Hamiltonian with respect to the transform 
Eq.~(\ref{eq_cantrans}) \cite{WXP13,WXBB15}
\begin{eqnarray}
\tauhat(\state,\gamma) & \equiv & 
\frac{\delta \Hhat(\state,\gamma)}{\delta \gamma}
\mbox{ and } \label{eq_tauhatdef} \\
\muAhat(\state,\gamma) & \equiv & 
\frac{\delta^2\Hhat(\state,\gamma)}{\delta \gamma^2} = \frac{\delta \tauhat(\state,\gamma)}{\delta \gamma}.  
\label{eq_muAhatdef}
\end{eqnarray}
For the differences of energy and shear stress caused by the transform this implies
\begin{eqnarray}
\delta \Hhat/V & \equiv & (\Hhat(\state,\gamma+\delta \gamma) - \Hhat(\state,\gamma))/V \nonumber \\
             & \approx & \tauhat(\state,\gamma) \ \delta \gamma + \frac{1}{2} \muAhat(\state,\gamma) \ \delta \gamma^2 
               \label{eq_dHaffine} \\
\delta \tauhat & \equiv  & \tauhat(\state,\gamma+\delta \gamma) - \tauhat(\state,\gamma) \nonumber \\
               & \approx & \muAhat(\state,\gamma) \ \delta \gamma.
\label{eq_dtauaffine}
\end{eqnarray}
With $\Hidhat(\state,\gamma)$ and $\Hexhat(\state,\gamma)$ being the standard kinetic and the 
(conservative) excess interaction contributions to the Hamiltonian 
$\Hhat(\state,\gamma) = \Hidhat(\state,\gamma) + \Hexhat(\state,\gamma)$,
this implies similar relations for the corresponding contributions
$\tauidhat$ and $\tauexhat$ to  $\tauhat =\tauidhat + \tauexhat$ and 
for the contributions $\muAidhat$ and $\muAexhat$ to $\muAhat = \muAidhat + \muAexhat$. 
For the ideal contributions this yields \cite{foot_monodisp,WXBB15,WKB15}
\begin{eqnarray}
\tauidhat(\state,\gamma) & = & - \frac{1}{V} \sum_{i=1}^N \vix \viy \label{eq_tauidhat} \mbox{ and } \\
\muAidhat(\state,\gamma) & = & \frac{1}{V} \sum_{i=1}^N \viy^2 \label{eq_muAidhat} 
\end{eqnarray}
where the minus sign for the shear stress is due to the minus sign in Eq.~(\ref{eq_cantrans}).
In this study we focus on pairwise additive excess energies $\Hexhat = \sum_l u(\rl)$
with $u(r)$ being a pair potential and where the 
running index $l$ labels the interaction between two particles $i < j$.
Straightforward application of the chain rule \cite{WXP13} shows that 
\begin{eqnarray}
\tauexhat(\state,\gamma) & = & 
\frac{1}{V} \sum_l \rl u^{\prime}(\rl) \ \nlx \nly   \label{eq_tauexhat} \ \mbox{ and } \\
\muAexhat(\state,\gamma) & = & \frac{1}{V} \sum_l  \left( \rl^2 u^{\prime\prime}(\rl)
- \rl u^{\prime}(\rl) \right) \nlx^2 \nly^2 \nonumber \\
& + & \frac{1}{V} \sum_l \rl u^{\prime}(\rl) \ \nly^2  \label{eq_muAexhat}
\end{eqnarray}
with $\rl$ being the distance between the beads and $\nvecl = \rvecl/\rl$ the normalized distance vector.
Note that Eq.~(\ref{eq_tauexhat}) is identical to the off-diagonal term 
of the standard Kirkwood stress tensor  \cite{AllenTildesleyBook}.

\section{Shear-stress fluctuation formula}
\label{theo_GF}
The stress-fluctuation formula Eq.~(\ref{eq_GF}) may be demonstrated elegantly \cite{WXP13,WKB15}
using the Lebowitz-Percus-Verlet transformations between conjugated ensembles \cite{Lebowitz67}
applied to the $\NVgT$- and $\NVtT$-ensembles. However, due to the steepest-descend approximation 
implict to this approach, which requires $\beta V\Geq \gg 1$, this approach can not be used for
transient networks since $\Geq(f>0)=0$.
We give here a more general demonstration of Eq.~(\ref{eq_GF}). 
The average equilibrium shear stress at a strain $\gamma$ is given by
\begin{equation}
\tau(\gamma) = \sum_{\state} \tauhat(\state,\gamma) \ \peq(\state,\gamma) \label{eq_tau_gam_peq}
\end{equation}
where the sum runs over all accessible states $\state$. 
The shear stress $\tauhat(\state,\gamma)$ of the state is given by Eq.~(\ref{eq_tauhatdef}) 
and the normalized equilibrium distribution $\peq(\state,\gamma)$ by
\begin{equation}
\peq(\state,\gamma)  = 
e^{-\beta \Hhat(\state,\gamma)} / \sum_{\state} e^{-\beta \Hhat(\state,\gamma)}.
\label{eq_peq_def}
\end{equation}
The task is now to compute the difference $\tau(\gamma+\delta \gamma) - \tau(\gamma)$
of the equilibrium shear stresses after and before the transform Eq.~(\ref{eq_cantrans}).
Using that
\begin{equation}
\exp[-\beta \Hhat(\state,\gamma+\delta \gamma)] \approx \exp(-\beta \Hhat(\state,\gamma)) \ (1-\beta \delta \Hhat)
\label{eq_expHhatexpand}
\end{equation}
with $\delta \Hhat$ being given by Eq.~(\ref{eq_dHaffine}) one shows that to leading order
the equilibrium distribution after the shear transformation may be expressed as
\begin{equation}
\frac{\peq(\state,\gamma+\delta \gamma)}{\peq(\state,\gamma)} \approx 
1-\beta \delta \Hhat + \beta \la \delta \Hhat \ra.
\label{eq_peq_expand}
\end{equation}
Using in addition Eq.~(\ref{eq_dtauaffine}) it is then readily seen that 
\begin{eqnarray}
\tau(\gamma+\delta \gamma) - \tau(\gamma) 
& \approx & \la \muAhat \ra \delta \gamma \label{eq_direct_two} \\
& - & \beta \left[ \la \tauhat(\state,\gamma) \delta \Hhat \ra 
- \la \tauhat(\state,\gamma) \ra \la \delta \Hhat \ra \right] \nonumber 
\end{eqnarray}
to leading order. Since according to Eq.~(\ref{eq_dHaffine}) we have 
$\delta \Hhat \approx V \tauhat(\state,\gamma) \ \delta \gamma$,
this leads to linear order in $\delta \gamma$ to 
\begin{eqnarray}
\frac{\tau(\gamma+\delta \gamma) - \tau(\gamma)}{\delta \gamma} & \approx &
\muA - \muFtild + \muFstar.\nonumber
\end{eqnarray}
We have thus confirmed Eq.~(\ref{eq_GF})
by only taking advantage of $\delta \gamma$ being arbitrarily small. This shows that Eq.~(\ref{eq_GF})
may also be used for liquids ($\Geq=0$) or for systems where $\beta V \Geq \ll 1$.
In the latter cases $\muA$ and $\muF$ simply become, respectively, identical or similar.

\section{Shear-stress relaxation}
\label{theo_Gt}

Following Ref.~\cite{DoiEdwardsBook} we present now an alternative demonstration 
of Eq.~(\ref{eq_key}) which does not require a finite equilibrium shear modulus. 
The time-dependent average shear stress $\tau(t)$ for $t > 0$ is given by
\begin{equation}
\tau(t) = \sum_{\state} \tauhat(\state,\gamma+\delta \gamma) \ p(t,\state) \label{eq_tau_gam}
\end{equation}
with $p(t,\state)$ being the time-dependent probability distribution of the state $\state$. 
We have $p(t=0,\state)=\peq(\state,\gamma)$ directly after the transformation at $t=0$
and $p(t,\state) \to \peq(\state,\gamma +\delta \gamma)$ for large times $t \gg \tstar$.
Consistently with Eq.~(\ref{eq_peq_expand}) it is useful here to expand the old equilibrium 
distribution in terms of the new one
\begin{equation}
\peq(\state,\gamma) 
\approx \peq(\state,\gamma+\delta \gamma) \left[1+\beta \delta \Hhat - \beta \la \delta \Hhat \ra\right].
\label{eq_peq_old2new}
\end{equation}
The time-dependent probability distribution is given by the general time evolution equation 
\cite{DoiEdwardsBook}
\begin{equation}
p(t,\state) = \sum_{\statep} G(\state,\statep;t-t') \ p(t'=0,\statep) \mbox{ for } t > 0
\label{eq_time_evolution}
\end{equation} 
with $G(\state,\statep;t-t')$ being an unspecified propagator of the system at $\gamma+\delta \gamma$.
We remind that a correlation function may be written as \cite{DoiEdwardsBook}
\begin{equation}
\la A(t) B(t') \ra = \sum_{\state,\statep} A(\state) G(\state,\statep;t-t') B(\statep) p(\statep,t').
\end{equation}
Inserting Eq.~(\ref{eq_time_evolution}) into Eq.~(\ref{eq_tau_gam}) and
using Eq.~(\ref{eq_peq_old2new}) this leads to
\begin{equation}
\tau(t) \approx \la \tauhat \ra
+ \beta \left(\la \tauhat(t) \delta \Hhat(t'=0) \ra - \la \tauhat \ra \la \delta \Hhat \ra \right)
\nonumber
\end{equation}
where all averages are computed using the final equilibrium distribution $\peq(\state,\gamma+\delta \gamma)$.
Substracting the reference shear stress before the transform 
$\tau(t=0^-) = \tau(\gamma)$ on both sides of the equation leads to
\begin{eqnarray}
\frac{\tau(t)-\tau(t=0^-)}{\delta \gamma} & \approx &
\frac{\tau(\gamma+\delta \gamma)-\tau(\gamma)}{\delta \gamma}  \\
& + & \beta V \left(\la \tauhat(t) \tauhat(0) \ra - \la \tauhat \ra^2 \right) \nonumber
\end{eqnarray}
to leading order. Taking finally $\delta \gamma \to 0$ and defining the ACF $c(t) \equiv \Ctild(t) - \muFstar$ 
this is equivalent to $G(t) = \Geq + c(t)$ in agreement with Ref.~\cite{WXB15}. 
Taking in addition advantage of the exact identity $h(t) = \Ctild(0) - \Ctild(t) = c(0) - c(t)$ 
\cite{DoiEdwardsBook}
relating the shear-stress ACF with the shear-stress MSD, this implies in turn Eq.~(\ref{eq_key}).

\section{Scaling with sampling time $\tsamp$}
\label{simu_dt}

\begin{figure}[t]
\centerline{\resizebox{1.0\columnwidth}{!}{\includegraphics*{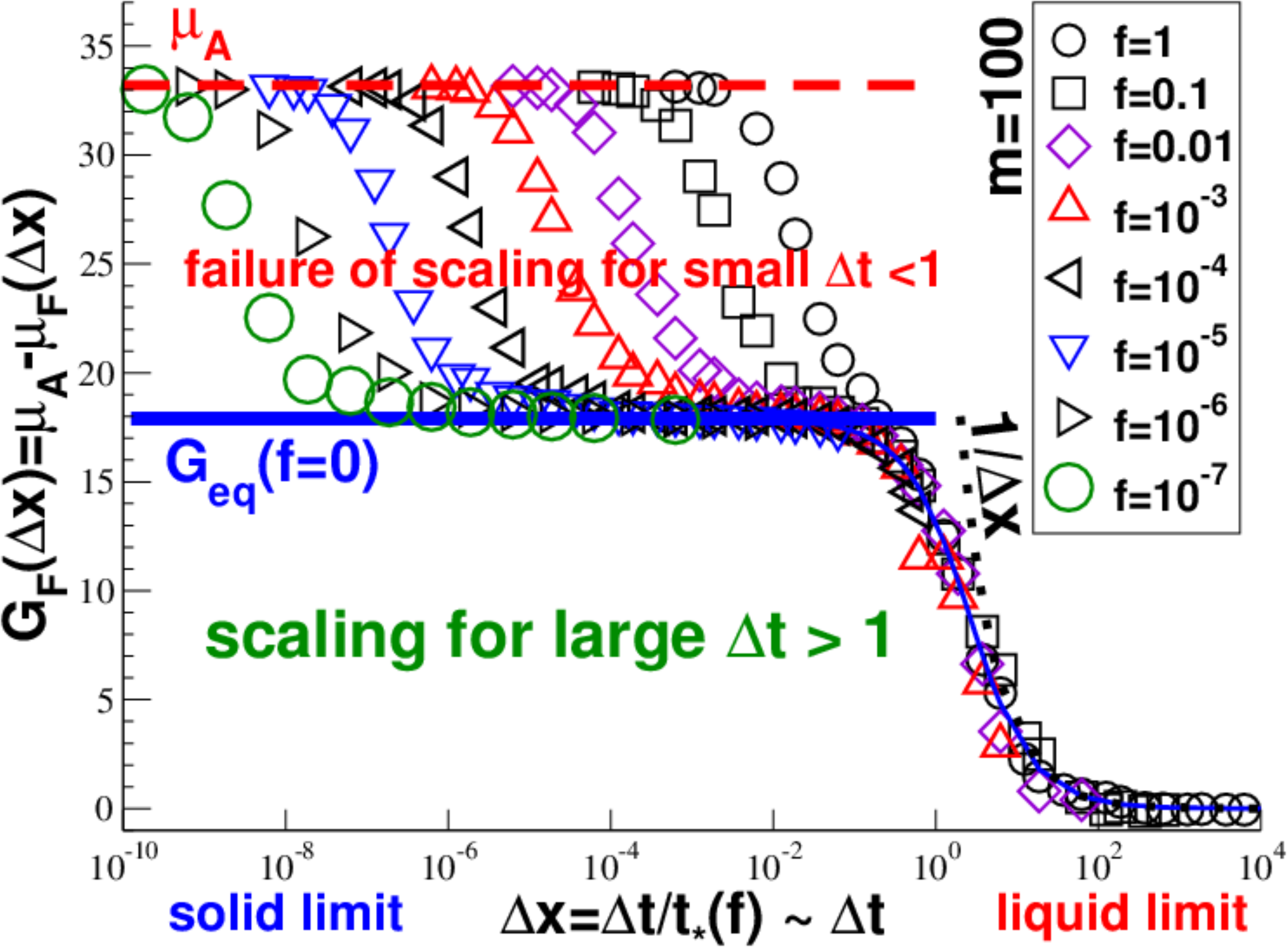}}}
\caption{$\GF(\tsamp)$ for subtrajectories of length $\tsamp \le \ttraj$ 
as a function of $\xsamp\equiv \tsamp/\tstar(f) \sim \tsamp$ for different $f$. The data scales for 
$\tsamp \gg 1$. The existence of an additional time scale is visible for small $\tsamp \ll 1$.
The thin solid line indicates Eq.~(\ref{eq_GFfDebye}). 
\label{fig_GF_dt}
}
\end{figure}

The dimensionless variable $\xsamp = \tsamp/\tstar$ has been changed in the main text 
only as a function of the attempt frequency $f$ while the sampling time $\tsamp$ was kept 
constant for clarity.
The scaling also holds if $\tsamp$ is varied at a constant terminal time $\tstar$ 
as was done for permanent networks \cite{WXB16}. 
As shown in Fig.~\ref{fig_GF_dt} for the stress-fluctuation formula $\GF$, this assumes 
that both $\tsamp$ and $\tstar(f)$ are sufficiently large.
The time average over a sampling time $\tsamp=\ttraj=10^{5}$ is replaced by averages over 
(independent) subintervals of length $\tsamp \le \ttraj$. Note that the largest values of $\xsamp$ 
indicated in Fig.~\ref{fig_GF_dt} for each $f$ correspond to the data given in 
Fig.~\ref{fig_static_f}. As expected, all data points collapse on a master 
curve (thin solid line) as long as $\xsamp$ remains sufficiently large.
The data for small $\xsamp$, where the scaling fails, correspond to $\tsamp \ll 1$. 
This merely shows that the additional time scale $\tauA$ (Fig.~\ref{fig_MSD_f}) becomes relevant.
Since $\muF$ vanishes for small $\tsamp$, this leads to the limit $\GF \to \muA$ 
indicated by the dashed line. 
\begin{figure}[t]
\centerline{\resizebox{1.0\columnwidth}{!}{\includegraphics*{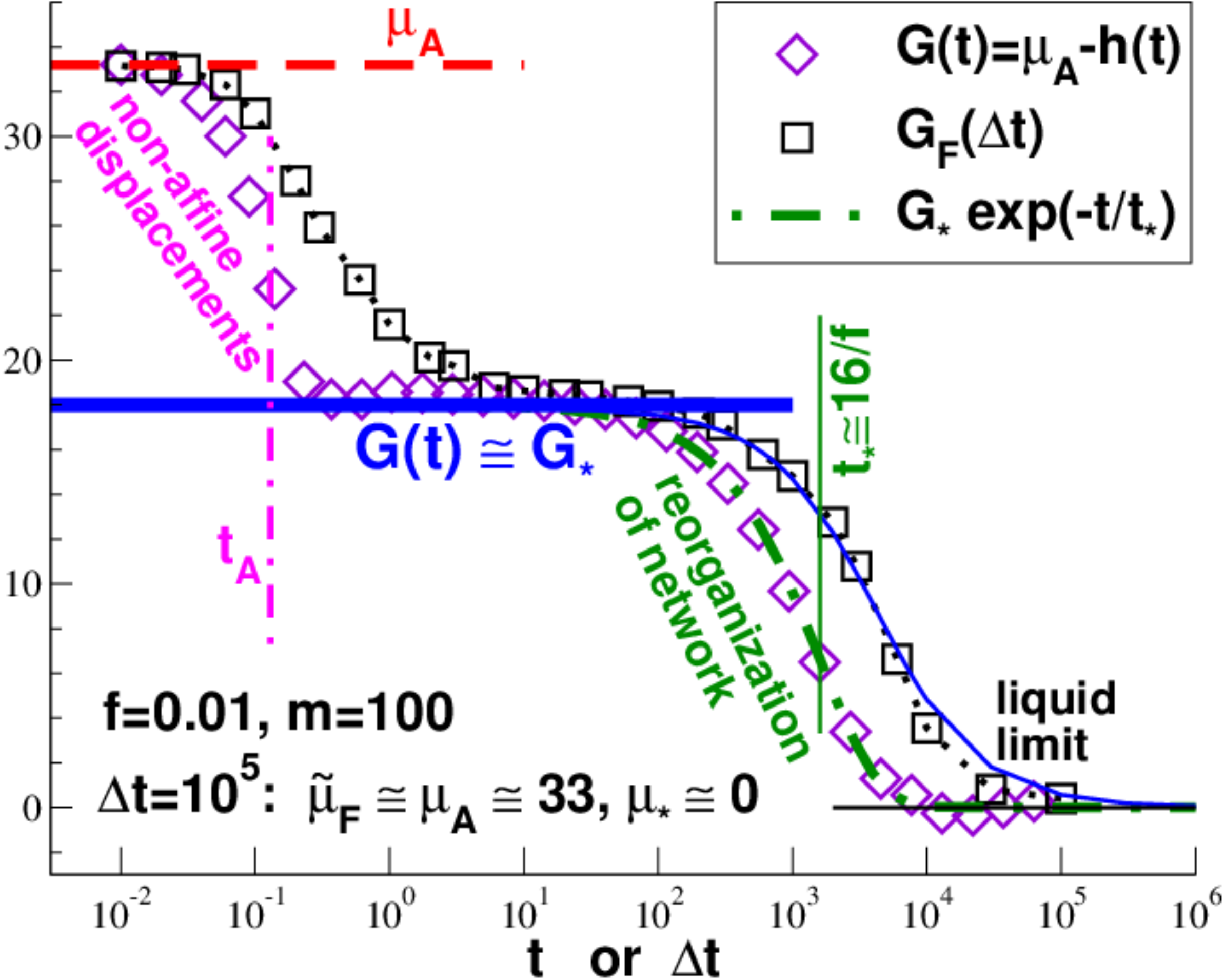}}}
\caption{Comparison of $G(t) = \muA - h(t)$ and $\GF(\tsamp) \equiv \muA - \muF(\tsamp)$ for 
one example in the liquid limit ($f=0.01$, $\xsamp=62.5$). 
Confirming Eq.~(\ref{eq_GF_Gt}), $\GF(\tsamp)$ is equivalent to the weighted integral over $G(t)$
indicated by the dotted line. Note that $G(t \approx \tsamp) \approx \GF(\tsamp)$ in the three 
time regimes where the response modulus has a plateau (shoulder). 
$\GF(\tsamp)$ is delayed with respect to $G(t)$
due to the strong weight of small times to the integral Eq.~(\ref{eq_Tver_def}).
The thin solid line indicates Eq.~(\ref{eq_GFfDebye}).
\label{fig_Gt_fmid}
}
\end{figure}

\section{Comparison of $G(t)$ and $\GF(\tsamp)$} 
\label{simu_conseq}
Assuming $y(t)$ to be an arbitrary well-behaved function of $t$ let us consider the linear functional
\begin{equation}
\Tver[y(t)] \equiv \frac{2}{\tsamp^2} \int_0^{\tsamp} \ddiff t \ (\tsamp-t) \ y(t)
\label{eq_Tver_def}
\end{equation}
motivated by Eq.~(\ref{eq_GF_Gt}). Note that contributions at the lower boundary of the 
integral have a strong weight due to the factor $(\tsamp-t)$ and that for a constant function
\begin{equation}
y(t)=c \mbox{ we have } \Tver[c] = c,
\label{eq_constfunc}
\end{equation}
i.e. the $\tsamp$-dependence drops out.
This does even hold to leading order 
if $y(t) \approx c$ only for large $t$ or for a finite $t$-window if
this window is sufficiently large. 
Assuming time translational invariance the shear stress fluctuation $\muF(\tsamp)$ is quite generally given by
$\muF(\tsamp) = \Tver[h(t)]$ \cite{WXB15,foot_polymer}.
Since $\muA$  is constant, Eq.~(\ref{eq_constfunc}) and Eq.~(\ref{eq_key}) imply
\begin{equation}
\GF(\tsamp) \equiv \muA - \muF(\tsamp) = \Tver[G(t)]
\label{eq_GFtsamp}
\end{equation}
in agreement with Eq.~(\ref{eq_GF_Gt}). 
According to Eq.~(\ref{eq_constfunc}), $\GF(\tsamp)$ should become similar to $G(t \approx \tsamp)$ 
in the three time windows $t \ll \tauA$, $\tauA \ll t \ll \tstar(f)$ and $\tstar(f) \ll t$ 
where $h(t)$ and $G(t)$ become approximatively constant (Fig.~\ref{fig_MSD_f}).
This is consistent with the data presented in Fig.~\ref{fig_Gt_fmid} for $f=0.01$. 
Note that specifically $\GF(\tsamp) \approx \muA$ for $\tsamp \ll \tauA$ in agreement 
with Fig.~\ref{fig_GF_dt}.
%



\newpage
\clearpage
\begin{thebibliography}{46}
\expandafter\ifx\csname natexlab\endcsname\relax\def\natexlab#1{#1}\fi
\expandafter\ifx\csname bibnamefont\endcsname\relax
  \def\bibnamefont#1{#1}\fi
\expandafter\ifx\csname bibfnamefont\endcsname\relax
  \def\bibfnamefont#1{#1}\fi
\expandafter\ifx\csname citenamefont\endcsname\relax
  \def\citenamefont#1{#1}\fi
\expandafter\ifx\csname url\endcsname\relax
  \def\url#1{\texttt{#1}}\fi
\expandafter\ifx\csname urlprefix\endcsname\relax\def\urlprefix{URL }\fi
\providecommand{\bibinfo}[2]{#2}
\providecommand{\eprint}[2][]{\url{#2}}

\bibitem[{\citenamefont{Alexander}(1998)}]{Alexander98}
\bibinfo{author}{\bibfnamefont{S.}~\bibnamefont{Alexander}},
  \bibinfo{journal}{Physics Reports} \textbf{\bibinfo{volume}{296}},
  \bibinfo{pages}{65 } (\bibinfo{year}{1998}).

\bibitem[{\citenamefont{Hansen and McDonald}(2006)}]{HansenBook}
\bibinfo{author}{\bibfnamefont{J.}~\bibnamefont{Hansen}} \bibnamefont{and}
  \bibinfo{author}{\bibfnamefont{I.}~\bibnamefont{McDonald}},
  \emph{\bibinfo{title}{Theory of simple liquids}}
  (\bibinfo{publisher}{Academic Press}, \bibinfo{address}{New York},
  \bibinfo{year}{2006}), \bibinfo{note}{3nd edition}.

\bibitem[{\citenamefont{G\"otze}(2009)}]{GoetzeBook}
\bibinfo{author}{\bibfnamefont{W.}~\bibnamefont{G\"otze}},
  \emph{\bibinfo{title}{Complex Dynamics of Glass-Forming Liquids: A
  Mode-Coupling Theory}} (\bibinfo{publisher}{Oxford University Press, Oxford},
  \bibinfo{year}{2009}).

\bibitem[{\citenamefont{de~Gennes}(1979)}]{DegennesBook}
\bibinfo{author}{\bibfnamefont{P.~G.} \bibnamefont{de~Gennes}},
  \emph{\bibinfo{title}{Scaling Concepts in Polymer Physics}}
  (\bibinfo{publisher}{Cornell University Press}, \bibinfo{address}{Ithaca, New
  York}, \bibinfo{year}{1979}).

\bibitem[{\citenamefont{Doi and Edwards}(1986)}]{DoiEdwardsBook}
\bibinfo{author}{\bibfnamefont{M.}~\bibnamefont{Doi}} \bibnamefont{and}
  \bibinfo{author}{\bibfnamefont{S.~F.} \bibnamefont{Edwards}},
  \emph{\bibinfo{title}{The Theory of Polymer Dynamics}}
  (\bibinfo{publisher}{Clarendon Press}, \bibinfo{address}{Oxford},
  \bibinfo{year}{1986}).

\bibitem[{\citenamefont{Witten and Pincus}(2004)}]{WittenPincusBook}
\bibinfo{author}{\bibfnamefont{T.}~\bibnamefont{Witten}} \bibnamefont{and}
  \bibinfo{author}{\bibfnamefont{P.~A.} \bibnamefont{Pincus}},
  \emph{\bibinfo{title}{Structured Fluids: Polymers, Colloids, Surfactants}}
  (\bibinfo{publisher}{Oxford University Press}, \bibinfo{address}{Oxford},
  \bibinfo{year}{2004}).

\bibitem[{\citenamefont{Rubinstein and Colby}(2003)}]{RubinsteinBook}
\bibinfo{author}{\bibfnamefont{M.}~\bibnamefont{Rubinstein}} \bibnamefont{and}
  \bibinfo{author}{\bibfnamefont{R.}~\bibnamefont{Colby}},
  \emph{\bibinfo{title}{Polymer Physics}} (\bibinfo{publisher}{Oxford
  University Press}, \bibinfo{address}{Oxford}, \bibinfo{year}{2003}).

\bibitem[{\citenamefont{Stauffer and Aharnony}(1994)}]{StaufferBook}
\bibinfo{author}{\bibfnamefont{D.}~\bibnamefont{Stauffer}} \bibnamefont{and}
  \bibinfo{author}{\bibfnamefont{A.}~\bibnamefont{Aharnony}},
  \emph{\bibinfo{title}{Introduction to percolation theory}}
  (\bibinfo{publisher}{Taylor \& Francis}, \bibinfo{address}{London},
  \bibinfo{year}{1994}).

\bibitem[{\citenamefont{Ulrich et~al.}(2006)\citenamefont{Ulrich, Mao,
  Goldbart, and Zippelius}}]{Zippelius06}
\bibinfo{author}{\bibfnamefont{S.}~\bibnamefont{Ulrich}},
  \bibinfo{author}{\bibfnamefont{X.}~\bibnamefont{Mao}},
  \bibinfo{author}{\bibfnamefont{P.}~\bibnamefont{Goldbart}}, \bibnamefont{and}
  \bibinfo{author}{\bibfnamefont{A.}~\bibnamefont{Zippelius}},
  \bibinfo{journal}{Europhysics Lett.} \textbf{\bibinfo{volume}{76}},
  \bibinfo{pages}{677} (\bibinfo{year}{2006}).

\bibitem[{\citenamefont{Wittmer et~al.}(2016)\citenamefont{Wittmer, Xu, and
  Baschnagel}}]{WXB16}
\bibinfo{author}{\bibfnamefont{J.~P.} \bibnamefont{Wittmer}},
  \bibinfo{author}{\bibfnamefont{H.}~\bibnamefont{Xu}}, \bibnamefont{and}
  \bibinfo{author}{\bibfnamefont{J.}~\bibnamefont{Baschnagel}},
  \bibinfo{journal}{Phys. Rev. E} \textbf{\bibinfo{volume}{93}},
  \bibinfo{pages}{012103} (\bibinfo{year}{2016}).

\bibitem[{\citenamefont{Wittmer et~al.}(2013)\citenamefont{Wittmer, Xu,
  Poli\'nska, Weysser, and Baschnagel}}]{WXP13}
\bibinfo{author}{\bibfnamefont{J.~P.} \bibnamefont{Wittmer}},
  \bibinfo{author}{\bibfnamefont{H.}~\bibnamefont{Xu}},
  \bibinfo{author}{\bibfnamefont{P.}~\bibnamefont{Poli\'nska}},
  \bibinfo{author}{\bibfnamefont{F.}~\bibnamefont{Weysser}}, \bibnamefont{and}
  \bibinfo{author}{\bibfnamefont{J.}~\bibnamefont{Baschnagel}},
  \bibinfo{journal}{J. Chem. Phys.} \textbf{\bibinfo{volume}{138}},
  \bibinfo{pages}{12A533} (\bibinfo{year}{2013}).

\bibitem[{\citenamefont{Wittmer
  et~al.}(2015{\natexlab{a}})\citenamefont{Wittmer, Xu, and
  Baschnagel}}]{WXB15}
\bibinfo{author}{\bibfnamefont{J.~P.} \bibnamefont{Wittmer}},
  \bibinfo{author}{\bibfnamefont{H.}~\bibnamefont{Xu}}, \bibnamefont{and}
  \bibinfo{author}{\bibfnamefont{J.}~\bibnamefont{Baschnagel}},
  \bibinfo{journal}{Phys. Rev. E} \textbf{\bibinfo{volume}{91}},
  \bibinfo{pages}{022107} (\bibinfo{year}{2015}{\natexlab{a}}).

\bibitem[{\citenamefont{Wittmer
  et~al.}(2015{\natexlab{b}})\citenamefont{Wittmer, Xu, Benzerara, and
  Baschnagel}}]{WXBB15}
\bibinfo{author}{\bibfnamefont{J.~P.} \bibnamefont{Wittmer}},
  \bibinfo{author}{\bibfnamefont{H.}~\bibnamefont{Xu}},
  \bibinfo{author}{\bibfnamefont{O.}~\bibnamefont{Benzerara}},
  \bibnamefont{and}
  \bibinfo{author}{\bibfnamefont{J.}~\bibnamefont{Baschnagel}},
  \bibinfo{journal}{Mol. Phys.} \textbf{\bibinfo{volume}{113}},
  \bibinfo{pages}{2881} (\bibinfo{year}{2015}{\natexlab{b}}).

\bibitem[{\citenamefont{Wittmer
  et~al.}(2015{\natexlab{c}})\citenamefont{Wittmer, Kriuchevskyi, Baschnagel,
  and Xu}}]{WKB15}
\bibinfo{author}{\bibfnamefont{J.~P.} \bibnamefont{Wittmer}},
  \bibinfo{author}{\bibfnamefont{I.}~\bibnamefont{Kriuchevskyi}},
  \bibinfo{author}{\bibfnamefont{J.}~\bibnamefont{Baschnagel}},
  \bibnamefont{and} \bibinfo{author}{\bibfnamefont{H.}~\bibnamefont{Xu}},
  \bibinfo{journal}{Eur. Phys. J. B} \textbf{\bibinfo{volume}{88}},
  \bibinfo{pages}{242} (\bibinfo{year}{2015}{\natexlab{c}}).

\bibitem[{\citenamefont{Allen and Tildesley}(1994)}]{AllenTildesleyBook}
\bibinfo{author}{\bibfnamefont{M.}~\bibnamefont{Allen}} \bibnamefont{and}
  \bibinfo{author}{\bibfnamefont{D.}~\bibnamefont{Tildesley}},
  \emph{\bibinfo{title}{Computer Simulation of Liquids}}
  (\bibinfo{publisher}{Oxford University Press}, \bibinfo{address}{Oxford},
  \bibinfo{year}{1994}).

\bibitem[{\citenamefont{Lebowitz et~al.}(1967)\citenamefont{Lebowitz, Percus,
  and Verlet}}]{Lebowitz67}
\bibinfo{author}{\bibfnamefont{J.~L.} \bibnamefont{Lebowitz}},
  \bibinfo{author}{\bibfnamefont{J.~K.} \bibnamefont{Percus}},
  \bibnamefont{and} \bibinfo{author}{\bibfnamefont{L.}~\bibnamefont{Verlet}},
  \bibinfo{journal}{Phys. Rev.} \textbf{\bibinfo{volume}{153}},
  \bibinfo{pages}{250} (\bibinfo{year}{1967}).

\bibitem[{\citenamefont{Barrat et~al.}(1988)\citenamefont{Barrat, Roux, Hansen,
  and Klein}}]{Barrat88}
\bibinfo{author}{\bibfnamefont{J.-L.} \bibnamefont{Barrat}},
  \bibinfo{author}{\bibfnamefont{J.-N.} \bibnamefont{Roux}},
  \bibinfo{author}{\bibfnamefont{J.-P.} \bibnamefont{Hansen}},
  \bibnamefont{and} \bibinfo{author}{\bibfnamefont{M.~L.} \bibnamefont{Klein}},
  \bibinfo{journal}{Europhys. Lett.} \textbf{\bibinfo{volume}{7}},
  \bibinfo{pages}{707} (\bibinfo{year}{1988}).

\bibitem[{\citenamefont{Wittmer et~al.}(2002)\citenamefont{Wittmer, Tanguy,
  Barrat, and Lewis}}]{WTBL02}
\bibinfo{author}{\bibfnamefont{J.~P.} \bibnamefont{Wittmer}},
  \bibinfo{author}{\bibfnamefont{A.}~\bibnamefont{Tanguy}},
  \bibinfo{author}{\bibfnamefont{J.-L.} \bibnamefont{Barrat}},
  \bibnamefont{and} \bibinfo{author}{\bibfnamefont{L.}~\bibnamefont{Lewis}},
  \bibinfo{journal}{Europhys. Lett.} \textbf{\bibinfo{volume}{57}},
  \bibinfo{pages}{423} (\bibinfo{year}{2002}).

\bibitem[{\citenamefont{Tanguy et~al.}(2002)\citenamefont{Tanguy, Wittmer,
  Leonforte, and Barrat}}]{TWLB02}
\bibinfo{author}{\bibfnamefont{A.}~\bibnamefont{Tanguy}},
  \bibinfo{author}{\bibfnamefont{J.~P.} \bibnamefont{Wittmer}},
  \bibinfo{author}{\bibfnamefont{F.}~\bibnamefont{Leonforte}},
  \bibnamefont{and} \bibinfo{author}{\bibfnamefont{J.-L.}
  \bibnamefont{Barrat}}, \bibinfo{journal}{Phys. Rev. B}
  \textbf{\bibinfo{volume}{66}}, \bibinfo{pages}{174205}
  (\bibinfo{year}{2002}).

\bibitem[{\citenamefont{Flenner and Szamel}(2015)}]{Szamel15}
\bibinfo{author}{\bibfnamefont{E.}~\bibnamefont{Flenner}} \bibnamefont{and}
  \bibinfo{author}{\bibfnamefont{G.}~\bibnamefont{Szamel}},
  \bibinfo{journal}{Phys. Rev. Lett.} \textbf{\bibinfo{volume}{107}},
  \bibinfo{pages}{105505} (\bibinfo{year}{2015}).

\bibitem[{\citenamefont{Xu et~al.}(2012)\citenamefont{Xu, Wittmer,
  Poli{\'n}ska, and Baschnagel}}]{XWP12}
\bibinfo{author}{\bibfnamefont{H.}~\bibnamefont{Xu}},
  \bibinfo{author}{\bibfnamefont{J.}~\bibnamefont{Wittmer}},
  \bibinfo{author}{\bibfnamefont{P.}~\bibnamefont{Poli{\'n}ska}},
  \bibnamefont{and}
  \bibinfo{author}{\bibfnamefont{J.}~\bibnamefont{Baschnagel}},
  \bibinfo{journal}{Phys. Rev. E} \textbf{\bibinfo{volume}{86}},
  \bibinfo{pages}{046705} (\bibinfo{year}{2012}).

\bibitem[{\citenamefont{Squire et~al.}(1969)\citenamefont{Squire, Holt, and
  Hoover}}]{Hoover69}
\bibinfo{author}{\bibfnamefont{D.~R.} \bibnamefont{Squire}},
  \bibinfo{author}{\bibfnamefont{A.~C.} \bibnamefont{Holt}}, \bibnamefont{and}
  \bibinfo{author}{\bibfnamefont{W.~G.} \bibnamefont{Hoover}},
  \bibinfo{journal}{Physica} \textbf{\bibinfo{volume}{42}},
  \bibinfo{pages}{388} (\bibinfo{year}{1969}).

\bibitem[{\citenamefont{Lutsko}(1989)}]{Lutsko89}
\bibinfo{author}{\bibfnamefont{J.~F.} \bibnamefont{Lutsko}},
  \bibinfo{journal}{J. Appl. Phys} \textbf{\bibinfo{volume}{65}},
  \bibinfo{pages}{2991} (\bibinfo{year}{1989}).

\bibitem[{\citenamefont{Mizuno et~al.}(2013)\citenamefont{Mizuno, Mossa, and
  Barrat}}]{Barrat13}
\bibinfo{author}{\bibfnamefont{H.}~\bibnamefont{Mizuno}},
  \bibinfo{author}{\bibfnamefont{S.}~\bibnamefont{Mossa}}, \bibnamefont{and}
  \bibinfo{author}{\bibfnamefont{J.-L.} \bibnamefont{Barrat}},
  \bibinfo{journal}{Phys. Rev. E} \textbf{\bibinfo{volume}{87}},
  \bibinfo{pages}{042306} (\bibinfo{year}{2013}).

\bibitem[{foo({\natexlab{a}})}]{foot_polymer}
\bibinfo{note}{A similar relation exists in polymer theory
  \cite{DoiEdwardsBook} expressing the radius of gyration of a chain as a
  weighted integral over internal mean-squared segment sizes.}

\bibitem[{foo({\natexlab{b}})}]{foot_simrel}
\bibinfo{note}{One may see Eq.~(\ref{eq_GF_Gt}) as the fundamental definition
  of the $\tsamp$-dependent stress-fluctuation formula. Similar expressions can
  be formulated for other response functions and associated stress-fluctuation
  formulae.}

\bibitem[{\citenamefont{Zilman et~al.}(2003)\citenamefont{Zilman, Kieffer,
  Molino, Porte, and Safran}}]{Porte03}
\bibinfo{author}{\bibfnamefont{A.}~\bibnamefont{Zilman}},
  \bibinfo{author}{\bibfnamefont{J.}~\bibnamefont{Kieffer}},
  \bibinfo{author}{\bibfnamefont{F.}~\bibnamefont{Molino}},
  \bibinfo{author}{\bibfnamefont{G.}~\bibnamefont{Porte}}, \bibnamefont{and}
  \bibinfo{author}{\bibfnamefont{S.~A.} \bibnamefont{Safran}},
  \bibinfo{journal}{Phys. Rev. Lett.} \textbf{\bibinfo{volume}{91}},
  \bibinfo{pages}{2003} (\bibinfo{year}{2003}).

\bibitem[{\citenamefont{Hed and Safran}(2006)}]{Safran06}
\bibinfo{author}{\bibfnamefont{G.}~\bibnamefont{Hed}} \bibnamefont{and}
  \bibinfo{author}{\bibfnamefont{S.}~\bibnamefont{Safran}},
  \bibinfo{journal}{Eur. Phys. J. E} \textbf{\bibinfo{volume}{19}},
  \bibinfo{pages}{69} (\bibinfo{year}{2006}).

\bibitem[{\citenamefont{Testard et~al.}(2008)\citenamefont{Testard, Oberdisse,
  and Ligoure}}]{Ligoure08}
\bibinfo{author}{\bibfnamefont{V.}~\bibnamefont{Testard}},
  \bibinfo{author}{\bibfnamefont{J.}~\bibnamefont{Oberdisse}},
  \bibnamefont{and} \bibinfo{author}{\bibfnamefont{C.}~\bibnamefont{Ligoure}},
  \bibinfo{journal}{Macromolecules} \textbf{\bibinfo{volume}{41}},
  \bibinfo{pages}{7219} (\bibinfo{year}{2008}).

\bibitem[{\citenamefont{Montarnal et~al.}(2011)\citenamefont{Montarnal,
  Capelot, Tournilhac, and Leibler}}]{Leibler11}
\bibinfo{author}{\bibfnamefont{D.}~\bibnamefont{Montarnal}},
  \bibinfo{author}{\bibfnamefont{M.}~\bibnamefont{Capelot}},
  \bibinfo{author}{\bibfnamefont{F.}~\bibnamefont{Tournilhac}},
  \bibnamefont{and} \bibinfo{author}{\bibfnamefont{L.}~\bibnamefont{Leibler}},
  \bibinfo{journal}{Science} \textbf{\bibinfo{volume}{334}},
  \bibinfo{pages}{965} (\bibinfo{year}{2011}).

\bibitem[{\citenamefont{Smallenburg et~al.}(2013)\citenamefont{Smallenburg,
  Leibler, and Sciortino}}]{Leibler13}
\bibinfo{author}{\bibfnamefont{F.}~\bibnamefont{Smallenburg}},
  \bibinfo{author}{\bibfnamefont{L.}~\bibnamefont{Leibler}}, \bibnamefont{and}
  \bibinfo{author}{\bibfnamefont{F.}~\bibnamefont{Sciortino}},
  \bibinfo{journal}{Phys. Rev. Lett.} \textbf{\bibinfo{volume}{111}},
  \bibinfo{pages}{188002} (\bibinfo{year}{2013}).

\bibitem[{\citenamefont{Rold\'an-Vargas
  et~al.}(2013)\citenamefont{Rold\'an-Vargas, Smallenburg, Kob, and
  Sciortino}}]{Kob13}
\bibinfo{author}{\bibfnamefont{S.}~\bibnamefont{Rold\'an-Vargas}},
  \bibinfo{author}{\bibfnamefont{F.}~\bibnamefont{Smallenburg}},
  \bibinfo{author}{\bibfnamefont{W.}~\bibnamefont{Kob}}, \bibnamefont{and}
  \bibinfo{author}{\bibfnamefont{F.}~\bibnamefont{Sciortino}},
  \bibinfo{journal}{J. Chem. Phys.} \textbf{\bibinfo{volume}{139}},
  \bibinfo{pages}{244910} (\bibinfo{year}{2013}).

\bibitem[{\citenamefont{Tonhauser et~al.}(2010)\citenamefont{Tonhauser, Wilms,
  Korth, Frey, and Friedrich}}]{Friedrich10}
\bibinfo{author}{\bibfnamefont{C.}~\bibnamefont{Tonhauser}},
  \bibinfo{author}{\bibfnamefont{D.}~\bibnamefont{Wilms}},
  \bibinfo{author}{\bibfnamefont{Y.}~\bibnamefont{Korth}},
  \bibinfo{author}{\bibfnamefont{H.}~\bibnamefont{Frey}}, \bibnamefont{and}
  \bibinfo{author}{\bibfnamefont{C.}~\bibnamefont{Friedrich}},
  \bibinfo{journal}{Macromolecular Rapid Comm.} \textbf{\bibinfo{volume}{31}},
  \bibinfo{pages}{2127} (\bibinfo{year}{2010}).

\bibitem[{\citenamefont{Berthier et~al.}(2010)\citenamefont{Berthier, Flenner,
  Jacquin, and Szamel}}]{Berthier10}
\bibinfo{author}{\bibfnamefont{L.}~\bibnamefont{Berthier}},
  \bibinfo{author}{\bibfnamefont{E.}~\bibnamefont{Flenner}},
  \bibinfo{author}{\bibfnamefont{H.}~\bibnamefont{Jacquin}}, \bibnamefont{and}
  \bibinfo{author}{\bibfnamefont{G.}~\bibnamefont{Szamel}},
  \bibinfo{journal}{Phys. Rev. E} \textbf{\bibinfo{volume}{81}},
  \bibinfo{pages}{031505} (\bibinfo{year}{2010}).

\bibitem[{\citenamefont{Berthier et~al.}(2011)\citenamefont{Berthier, Jacquin,
  and Zamponi}}]{Berthier11a}
\bibinfo{author}{\bibfnamefont{L.}~\bibnamefont{Berthier}},
  \bibinfo{author}{\bibfnamefont{H.}~\bibnamefont{Jacquin}}, \bibnamefont{and}
  \bibinfo{author}{\bibfnamefont{F.}~\bibnamefont{Zamponi}},
  \bibinfo{journal}{Phys. Rev. E} \textbf{\bibinfo{volume}{84}},
  \bibinfo{pages}{051103} (\bibinfo{year}{2011}).

\bibitem[{\citenamefont{Wittmer et~al.}(1998)\citenamefont{Wittmer, Milchev,
  and Cates}}]{WMC98b}
\bibinfo{author}{\bibfnamefont{J.~P.} \bibnamefont{Wittmer}},
  \bibinfo{author}{\bibfnamefont{A.}~\bibnamefont{Milchev}}, \bibnamefont{and}
  \bibinfo{author}{\bibfnamefont{M.~E.} \bibnamefont{Cates}},
  \bibinfo{journal}{J. Chem. Phys.} \textbf{\bibinfo{volume}{109}},
  \bibinfo{pages}{834} (\bibinfo{year}{1998}).

\bibitem[{\citenamefont{Huang et~al.}(2006)\citenamefont{Huang, Xu, Crevel,
  Wittmer, and Ryckaert}}]{HXCWR06}
\bibinfo{author}{\bibfnamefont{C.~C.} \bibnamefont{Huang}},
  \bibinfo{author}{\bibfnamefont{H.}~\bibnamefont{Xu}},
  \bibinfo{author}{\bibfnamefont{F.}~\bibnamefont{Crevel}},
  \bibinfo{author}{\bibfnamefont{J.}~\bibnamefont{Wittmer}}, \bibnamefont{and}
  \bibinfo{author}{\bibfnamefont{J.-P.} \bibnamefont{Ryckaert}}, in
  \emph{\bibinfo{booktitle}{Computer Simulations in Condensed Matter: from
  Materials to Chemical Biology}} (\bibinfo{publisher}{Springer, Lect. Notes
  Phys., International School of Solid State Physics},
  \bibinfo{address}{Berlin/Heidelberg}, \bibinfo{year}{2006}), vol.
  \bibinfo{volume}{704}, pp. \bibinfo{pages}{379--418}.

\bibitem[{\citenamefont{Duering et~al.}(1991)\citenamefont{Duering, Kremer, and
  Grest}}]{DKG91}
\bibinfo{author}{\bibfnamefont{E.}~\bibnamefont{Duering}},
  \bibinfo{author}{\bibfnamefont{K.}~\bibnamefont{Kremer}}, \bibnamefont{and}
  \bibinfo{author}{\bibfnamefont{G.~S.} \bibnamefont{Grest}},
  \bibinfo{journal}{Phys. Rev. Lett.} \textbf{\bibinfo{volume}{67}},
  \bibinfo{pages}{3531} (\bibinfo{year}{1991}).

\bibitem[{\citenamefont{Klix et~al.}(2012)\citenamefont{Klix, Ebert, Weysser,
  Fuchs, Maret, and Keim}}]{Klix12}
\bibinfo{author}{\bibfnamefont{C.}~\bibnamefont{Klix}},
  \bibinfo{author}{\bibfnamefont{F.}~\bibnamefont{Ebert}},
  \bibinfo{author}{\bibfnamefont{F.}~\bibnamefont{Weysser}},
  \bibinfo{author}{\bibfnamefont{M.}~\bibnamefont{Fuchs}},
  \bibinfo{author}{\bibfnamefont{G.}~\bibnamefont{Maret}}, \bibnamefont{and}
  \bibinfo{author}{\bibfnamefont{P.}~\bibnamefont{Keim}},
  \bibinfo{journal}{Phys. Rev. Lett.} \textbf{\bibinfo{volume}{109}},
  \bibinfo{pages}{178301} (\bibinfo{year}{2012}).

\bibitem[{foo({\natexlab{c}})}]{foot_nsp}
\bibinfo{note}{This could be generalized by imposing an energy penalty
  $\Unsp(\nsp)$ with $\nsp$ being the number of springs per bead. One may,
  e.g., consider $\Unsp(\nsp)=0$ for $\nsp=0,1$ and $2$, $\Unsp(\nsp=3)=10$ and
  $\Unsp(\nsp)=\infty$ for all other $\nsp$ and in this way generate
  equilibrium polymer systems \cite{WMC98b,HXCWR06} with a few branching points
  where $\nsp=3$.}

\bibitem[{foo({\natexlab{d}})}]{foot_DB}
\bibinfo{note}{Detailed balance implies that a bond can neither be broken nor
  created with $r > \rc$. Note also that the neighbor list of beads around the
  pivot monomer $i$ contains exactly the same number of possible new sites
  before and after the hopping of the spring end from monomer $j$ to $k$
  sketched in panel (c) of Fig.~\ref{fig_sketch}, i.e. no additional weights
  are needed to ensure detailed balance \cite{LandauBinderBook}.}

\bibitem[{\citenamefont{Landau and Binder}(2000)}]{LandauBinderBook}
\bibinfo{author}{\bibfnamefont{D.~P.} \bibnamefont{Landau}} \bibnamefont{and}
  \bibinfo{author}{\bibfnamefont{K.}~\bibnamefont{Binder}},
  \emph{\bibinfo{title}{A Guide to Monte Carlo Simulations in Statistical
  Physics}} (\bibinfo{publisher}{Cambridge University Press},
  \bibinfo{address}{Cambridge}, \bibinfo{year}{2000}).

\bibitem[{foo({\natexlab{e}})}]{foot_powertwo}
\bibinfo{note}{Assuming time-reversal symmetry this power is expected
  \cite{WKB15,HansenBook}. Time-reversal symmetry applies on this time scale
  since the Langevin thermostat is irrelevant below a time of order $1/\zeta
  \approx 1 \gg \tauA$.}

\bibitem[{foo({\natexlab{f}})}]{foot_notdone}
\bibinfo{note}{Equation~(\ref{eq_dmuFtild_bound}) is merely stated here as a
  phenomenological description of the data. A theoretical justification will be
  presented elsewhere.}

\bibitem[{foo({\natexlab{g}})}]{foot_monodisp}
\bibinfo{note}{For convenience, we assume monodisperse particles of unit mass
  $m=1$, i.e. particle momenta $\pvec_i$ and velocities $\vvec_i$ are
  equivalent.}

\bibitem[{\citenamefont{Goldstein et~al.}(2001)\citenamefont{Goldstein, Safko,
  and Poole}}]{Goldstein}
\bibinfo{author}{\bibfnamefont{H.}~\bibnamefont{Goldstein}},
  \bibinfo{author}{\bibfnamefont{J.}~\bibnamefont{Safko}}, \bibnamefont{and}
  \bibinfo{author}{\bibfnamefont{C.}~\bibnamefont{Poole}},
  \emph{\bibinfo{title}{Classical Mechanics}}
  (\bibinfo{publisher}{Addison-Wesley}, \bibinfo{year}{2001}),
  \bibinfo{note}{3nd edition}.

\end{thebibliography}
\end{document}